\pdfoutput=1

\documentclass[preprint,3p,times]{elsarticle}

\usepackage{lineno,hyperref}
\usepackage{bm}
\usepackage{amsmath}
\modulolinenumbers[5]

\journal{Journal of Computational Physics}

\begin{document}

\begin{frontmatter}

\title{A new approach to wall modeling in LES of incompressible flow \\via function enrichment}

\author{B. Krank}
\ead{krank@lnm.mw.tum.de}
\author{W. A. Wall\corref{correspondingauthor1}}
\cortext[correspondingauthor1]{Corresponding author at: Institute for Computational Mechanics, Technische Universit\"at M\"unchen, Boltzmannstr. 15, 85748 Garching, Germany. Tel.: +49 89 28915300; fax: +49 89 28915301}
\ead{wall@lnm.mw.tum.de}

\address{Institute for Computational Mechanics, Technische Universit\"at M\"unchen,\\ Boltzmannstr. 15, 85748 Garching, Germany}

\begin{abstract}
A novel approach to wall modeling for the incompressible Navier-Stokes equations including flows of moderate and large Reynolds numbers is presented. The basic idea is that a problem-tailored function space allows prediction of turbulent boundary layer gradients with very coarse meshes. The proposed function space consists of a standard polynomial function space plus an enrichment, which is constructed using Spalding's law-of-the-wall. The enrichment function is not enforced but "allowed" in a consistent way and the overall methodology is much more general and also enables other enrichment functions. The proposed method is closely related to detached-eddy simulation as near-wall turbulence is modeled statistically and large eddies are resolved in the bulk flow. Interpreted in terms of a three-scale separation within the variational multiscale method, the standard scale resolves large eddies and the enrichment scale represents boundary layer turbulence in an averaged sense. The potential of the scheme is shown applying it to turbulent channel flow of friction Reynolds numbers from $Re_{\tau}=590$ and up to $5,000$, flow over periodic constrictions at the Reynolds numbers $Re_H=10,595$ and $19,000$ as well as backward-facing step flow at $Re_h=5,000$, all with extremely coarse meshes. Excellent agreement with experimental and DNS data is observed with the first grid point located at up to $y_1^+=500$ and especially under adverse pressure gradients as well as in separated flows.
\end{abstract}


\end{frontmatter}

%\linenumbers

\section{Introduction}

Large-eddy simulation (LES) becomes prohibitively expensive for moderate and high Reynolds numbers if near-wall turbulence is resolved. Grid-resolution requirements enabling prediction of the necessary scales depend on the friction Reynolds number approximately as $Re_{\tau}^2$ \cite{Baggett97}. Small computation cells in the boundary layer come along with severe constraints on the time step size to resolve the temporal scales of momentum-transfer mechanisms and to be compliant with the Courant-Friedrichs-Lewy (CFL) condition if explicit time integration schemes are utilized. 

The concept of wall modeling was therefore introduced in early works on LES of high-Reynolds-number flow by Deardorff \cite{Deardorff70} and Schumann \cite{Schumann75} in an attempt to circumvent the resolution dependence on wall units. Wall modeling implies that near-wall turbulence and the accompanying momentum transfer are not resolved in detail but modeled in a statistical sense. With near-wall turbulence modeled, the size of dominating eddies in the bulk of the flow are governed by geometrical scales of boundary conditions with resolution requirements increasing approximately as $Re^{0.4}$ with the Reynolds number of the bulk flow \cite{Piomelli02}. 

Common approaches in wall modeling focus on imposing synthetic boundary conditions prescribing tractions instead of no-slip velocity, see reviews e.g. in \cite{Cabot00,Piomelli02,Piomelli08}.  This comes along with the major advantage that the velocity gradient does not have to be resolved explicitly by the scheme, enabling very coarse meshes. Yet accurate models are required to predict the correct stresses. Simple equilibrium models allow for direct modeling of the tractions but are prone to inaccurate predictions in separated regions or flows with high pressure gradients \cite{Piomelli02}. More accurate two-layer models have been developed, e.g. solving the simplified thin-boundary-layer equations (TBLE) on a separate domain between the wall and the first off-wall node to predict the momentum transfer inside the boundary layer \cite{Balaras96,Wang02,ChenHickel14}. 
For a comprehensive overview of wall-layer models it is referred to Piomelli and Balaras \cite{Piomelli02} and Piomelli \cite{Piomelli08}. 
Alternative methods of imposing approximate boundary conditions have been proposed such as weak no-slip boundary conditions \cite{Bazilevs07b,Gamnitzer12} or the filtered-wall model \cite{Bhattacharya08} which are not frequently used but have lead to new insights in the field of wall modeling.

Hybrid RANS/LES methods including detached-eddy simulation (DES) represent another paradigm for wall modeling \cite{Spalart97,Frohlich08}. Instead of employing separate domains for near-wall and bulk turbulence such as in two-layer models, different subgrid models are applied in the respective regions of a single mesh. Reynolds-averaged Navier-Stokes equations are commonly employed in the wall region and LES subgrid closures in the bulk of the flow. The sharp velocity gradient present in high-Reynolds-number flow necessitates many computation points in wall-normal direction; under-resolution results in an imprecise prediction of the wall stresses leading to a log-layer mismatch \cite{Piomelli02}.

A major reason for DES requiring many grid points in wall-normal direction to be able to resolve the gradient accurately is the common application of low-order computational methods. In this work, we propose a problem-tailored high-order method in this region that is capable of resolving the velocity gradient with very coarse meshes. This is done by employing general knowledge about turbulent boundary layers without prescribing the velocity profile itself. Theoretical considerations on methods that allow for constructing customized numerical methods have first been introduced by Melenk and Babu\v{s}ka \cite{Melenk96} with their partition-of-unity method (PUM). Belytschko and Black \cite{Belytschko99} have subsequently suggested a formalism that allows for construction of a problem-tailored computational method in the application of crack propagation in solid mechanics. An enrichment function representing an approximate analytical solution is usually used to extend the solution space of the method, besides the standard polynomial function space. For a comprehensive overview of the method we refer to the review articles in \cite{Belytschko09,Fries10}. 

Applications of this method in the field of fluid mechanics can be found in several academic examples such as enrichment with analytical high-gradient solutions of the convection-diffusion equation \cite{Abbas10}, simulating a sharp corner in Stokes flow via an asymptotic expansion as enrichment \cite{Foucard14} or resolving the bottom boundary layer of oceanic flow via a logarithmic enrichment function applied to a 1D water column \cite{Hanert07}. A recent publication suggests enrichment with modes obtained by a proper orthogonal decomposition to resolve the boundary layer of a stochastically forced Burger's equation \cite{Chen14}. The general framework can also be used to resolve other features of the solution besides high gradients, such as jumps or kinks, and can even be used to cut elements (see, e.g., \cite{Schott14}). In general, the enrichment function is not prescribed as solution but the method "chooses" the best solution among all functions available in its function space in a consistent manner.

In this study, we suggest to extend the standard solution space with the law-of-the-wall due to Spalding as enrichment function such that the numerical method is able to represent the high-gradient velocity profile in a turbulent boundary layer with very coarse meshes. The idea follows the paradigm of DES as only the large eddies away from the wall are resolved explicitly and near-wall turbulent structures are represented in a statistical sense. The construction of the method does however not require blending of turbulence models as the approach separates the statistical model a priori from the eddy-resolving space according to the variational multiscale method. Yet it is required that subgrid turbulence is modeled accurately such that the computational method is able to find an appropriate solution. Therefore a multifractal subgrid scale model embedded in a variational multiscale method is applied as subgrid scale model which gives excellent results for wall-resolved LES \cite{Rasthofer12} and has been successfully extended to passive-scalar mixing \cite{Rasthofer14a} as well as low-Mach-number flow with variable density \cite{Rasthofer14b}.

%One important difference between our approach and most other applications of this method is that we apply a single empirical enrichment function that is widely recognized. In this respect, most other applications suggest several enrichment functions at the same time that often are obtained via simplifications of the original problem statement and mathematical relations, such as modes obtained by a proper orthogonal decomposition \cite{Chen14} or asymptotic expansions \cite{Foucard14}.

The present article is organized as follows. In Section \ref{sec:ns}, a weighted residual formulation of the incompressible Navier-Stokes equations is presented. Subsequently the enrichment of the function space with an appropriate gradient-capturing function is proposed in Section \ref{sec:enrichment} and methods are presented for adaptivity in space and time. Turbulence modeling in the framework of the variational multiscale method and necessary adaptations to the new function space are revisited in Section \ref{sec:subgrmodeling}. The method is validated with turbulent channel flow at moderate to moderately large Reynolds number, flow past periodic hills as well as flow over a backward-facing step in Section \ref{sec:numerical}. Conclusions close the article in Section \ref{sec:conclusions}.

\section{Incompressible Navier-Stokes equations}
\label{sec:ns}
The incompressible Navier-Stokes equations are considered in this work as outlined in the following. The weighted-residual formulation presented in the second subsection is the starting point for approximation of the solution spaces in Section \ref{sec:enrichment}.
\subsection{Problem statement}
The incompressible Navier-Stokes equations in convective form are considered as
\begin{equation}
\frac{\partial \bm{u}}{\partial t} + \bm{u} \cdot \nabla \bm{u} + \nabla p - 2 \nu \nabla \cdot \bm{\epsilon} (\bm{u}) = \bm{f} \text{ \hspace{0.5cm} in } \Omega \times (0,\mathcal{T})
\label{eq:mom}
\end{equation}
\begin{equation}
\nabla \cdot \bm{u} = 0  \text{ \hspace{0.5cm} in } \Omega \times (0,\mathcal{T})
\label{eq:conti}
\end{equation}
with the fluid velocity $\bm{u}=(u_1,u_2,u_3)^{T}$, the pressure $p$, the time t, the kinematic viscosity $\nu$ and the symmetric rate-of-deformation-tensor $\bm{\epsilon}(\bm{u})= \frac{1}{2}(\nabla \bm{u}+(\nabla \bm{u})^T)$. The body force vector is denoted $\bm{f}$, the spatial domain $\Omega$ and the simulation time $\mathcal{T}$. At $t=0$, a divergence-free initial velocity field is prescribed:
\begin{equation}
\bm{u}=\bm{u}_0  \text{ \hspace{0.5cm} in } \Omega \times \{0\}
\end{equation}
Dirichlet boundary conditions are defined as 
\begin{equation}
\bm{u} = \bm{u}_D  \text{ \hspace{0.5cm} on }  \Gamma_D \times (0,\mathcal{T})
\end{equation}
and traction boundary conditions are applied on the Neumann boundary
\begin{equation}
\bm{\sigma} \cdot \bm{n} = \bm{h}  \text{ \hspace{0.5cm} on }  \Gamma_N \times (0,\mathcal{T})
\label{eq:neum}
\end{equation}
where the Cauchy-Stress tensor is $\bm{\sigma} = - p \bm{I} + 2 \nu \bm{\epsilon}({\bm{u}})$. It is assumed that $\Gamma_D \cup \Gamma_N = \emptyset$ and $\Gamma_D \cap \Gamma_N = \Gamma$.

\subsection{Weighted residual formulation}

A weighted-residual formulation is obtained with a standard procedure by multiplying the momentum equation (\ref{eq:mom}) with a weighting function $\bm{v} \in \mathcal{V}_{\bm{v}}$ and the continuity equation (\ref{eq:conti}) with $q \in \mathcal{V}_{q}$. Appropriate spaces for $\bm{u} \in \mathcal{S}_{\bm{u}}$, $p \in \mathcal{S}_{p}$ as well as $\mathcal{V}_{\bm{v}}$ and $\mathcal{V}_{q}$ are assumed. The choice of the discrete solution and weighting function spaces $\mathcal{S}_{\bm{u}}^h$ and $\mathcal{V}_{\bm{v}}^h$ is the main innovation presented in this article and is discussed in the subsequent Section \ref{sec:enrichment}. The equations are integrated over the domain $\Omega$, the pressure and viscous terms are integrated by parts and the Neumann boundary conditions (\ref{eq:neum}) are applied to the arising boundary integrals. The result reads
\begin{equation}
\mathcal{B}_{NS}(\bm{v},q;\bm{u},p)=l(\bm{v})
\label{eq:weightedresidual}
\end{equation}
with the left hand side of the momentum equation and the contribution of the continuity equation
\begin{equation}
\mathcal{B}_{NS}(\bm{v},q;\bm{u},p)=(\bm{v},\frac{\partial \bm{u}}{\partial t}) + (\bm{v},\bm{u} \cdot \nabla \bm{u}) - (\nabla \cdot \bm{v}, p) + (\bm{\epsilon}(\bm{v}),2 \nu \bm{\epsilon} (\bm{u}))+(q,\nabla \cdot \bm{u})
\end{equation}
and the right hand side of the momentum equation
\begin{equation}
l(\bm{v})=(\bm{v},\bm{f})+(\bm{v},\bm{h})_{\Gamma_N}.
\end{equation}
The L2-inner product is defined as usual $(\cdot,\cdot)=(\cdot,\cdot)_{\Omega}$ and $(\cdot,\cdot)_{\Gamma_N}$ defines an integral over the Neumann boundary $\Gamma_N$.

\section{Capturing the boundary layer via function enrichment}
\label{sec:enrichment}
The computational method applied to solve the incompressible Navier-Stokes equations has large influence on the quality of the solution and number of cells required. The method presented in the following utilizes a problem-tailored solution space distinguishing drastically from standard methods. The solution space of the method is capable of resolving high boundary layer gradients and adapts to local characteristics of the flow. This is done by extending the solution space with the help of an approximate analytical representation of the mean velocity profile given as the law-of-the-wall. With a solution space capable of resolving the high gradient, the solution is not prescribed but the numerical method is able to find an appropriate solution in the offered function space, provided that turbulence is modeled accurately. The approach applied for turbulence modeling is discussed in Section \ref{sec:subgrmodeling}. As numerical method, a variant of the PUM or the extended finite element method (XFEM) is chosen as it provides a framework for constructing such a customized space.

\subsection{Enriching the solution space}
The extended finite element method suggests a solution space $\bm{u}^h(\bm{x},t)$ consisting of two contributions, a standard $\bar{\bm{u}}^h(\bm{x},t)$ and an enrichment part $\tilde{\bm{u}}^h(\bm{x},t)$ as
\begin{equation}
\bm{u}^h(\bm{x},t)=\bar{\bm{u}}^h(\bm{x},t)+\tilde{\bm{u}}^h(\bm{x},t)
\label{eq:space}
\end{equation}
dependent on the spatial location vector $\bm{x}=(x_1,x_2,x_3)^T$ and assuming direct sum decomposition of the underlying discrete solution spaces ${\mathcal{S}}_{\bm{u}}^h=\bar{\mathcal{S}}_{\bm{u}}^h\times \tilde{\mathcal{S}}_{\bm{u}}^h$, which are identified with a characteristic element length $h$.
The standard finite element expansion with shape functions $N_B^{\bm{u}}$ and degrees of freedom $\bar{\bm{u}}_B$ is
\begin{equation}
\bar{\bm{u}}^h(\bm{x},t)=\sum_{B \in N^{\bm{u}}} N_B^{\bm{u}}(\bm{x}) \bar{\bm{u}}_B.
\end{equation}
The enrichment including additional degrees of freedom $\tilde{\bm{u}}_{B}$ is with the same partition of unity $N_B^{\bm{u}}$ defined as
\begin{equation}
\tilde{\bm{u}}^h(\bm{x},t)=\sum_{B \in N_{enr}^{\bm{u}}} N_{B}^{\bm{u}}(\bm{x}) (\psi(\bm{x},t)-\psi(\bm{x}_{B},t)) r^h(\bm{x}) \tilde{\bm{u}}_{B} 
\label{eq:enrichment}
\end{equation}
where only a subset of nodes in the vicinity of the wall $N_{enr}^{\bm{u}} \subset N^{\bm{u}}$ is enriched. The enrichment function $\psi(\bm{x},t)$ represents a problem-tailored profile, e.g. an analytical or approximate solution of the underlying problem and the enrichment function suggested in this article is presented in the following subsection. Subtracting the enrichment function by its nodal values $\psi(\bm{x}_{B},t)$ yields zero on the nodes facilitating post processing and application of boundary conditions. 

\begin{figure}
\center
\setlength{\unitlength}{1mm}
\begin{picture}(70,30)
\put(5,5){\vector(1,0){57}}
\put(5,5){\vector(0,1){15}}

\thicklines
\put(5,5){\circle*{1.5}}
\put(15,5){\circle*{1.5}}
\put(25,5){\circle*{1.5}}
\put(35,5){\circle*{1.5}}
\put(45,5){\circle*{1.5}}
\put(55,5){\circle*{1.5}}

\put(6,4){\line(0,1){2}}
\put(4,4){\line(0,1){2}}
\put(4,6){\line(1,0){2}}
\put(4,4){\line(1,0){2}}

\put(16,4){\line(0,1){2}}
\put(14,4){\line(0,1){2}}
\put(14,6){\line(1,0){2}}
\put(14,4){\line(1,0){2}}

\put(26,4){\line(0,1){2}}
\put(24,4){\line(0,1){2}}
\put(24,6){\line(1,0){2}}
\put(24,4){\line(1,0){2}}

\put(36,4){\line(0,1){2}}
\put(34,4){\line(0,1){2}}
\put(34,6){\line(1,0){2}}
\put(34,4){\line(1,0){2}}

\put(7,18){$r^h(\bm{x})$}
\put(59,1){$y$}
\put(2,14){$1$}
\put(2,1){$y=0$}

\put(45,20){\circle*{1.5}}
\put(46,14){\line(0,1){2}}
\put(44,14){\line(0,1){2}}
\put(44,16){\line(1,0){2}}
\put(44,14){\line(1,0){2}}

\put(47.5,19.2){\footnotesize \text{standard FE node}}
\put(47.5,14.2){\footnotesize \text{enrichment node}}
\thinlines
\put(4,15){\line(1,0){1}}
\thicklines
\put(5,15){\line(1,0){20}}
\put(25,15){\line(1,-1){10}}

\end{picture}
\caption{Ramp function for blending.}
\label{fig:ramp}
\end{figure}
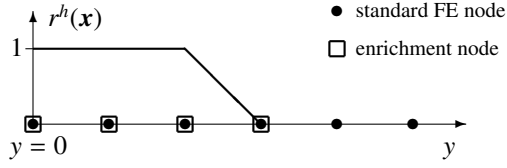

Special treatment of the enrichment in the blending area is required. On the interface towards non-enriched elements, the enrichment does not vanish yielding Neumann boundary conditions with $\bm{h}=\bm{0}$ on the enrichment nodes. As these nodes are subject both to in- and outflow of the domain and it is well-known that Neumann boundary conditions are ill-posed as inflow boundary for the incompressible Navier-Stokes equations, convergence problems are observed. This problem is circumvented by multiplying the enrichment with a ramp function $r^h(\bm{x})$ as detailed in \cite{Fries08} and depicted in figure \ref{fig:ramp}.

In a turbulent boundary layer, high-gradient solutions are only obtained for the velocity profile. The pressure space therefore remains unaltered and is given as a standard finite element expansion
\begin{equation}
p^h(\bm{x},t)=\sum_{B \in N^{p}} N_B^p(\bm{x}) p_B.
\end{equation}

As partition of unity we choose shape functions of a standard eight-noded trilinearly interpolated hexahedral finite element throughout this article for $N^{\bm{u}}$ and $N^{p}$. Any other Lagrangian finite element could be employed as well including higher-order elements and unstructured grids.

%figure showing ramp function

\subsection{Law-of-the-wall as enrichment}
\label{sec:lotw}
\begin{figure}
\center
\includegraphics[scale=0.6]{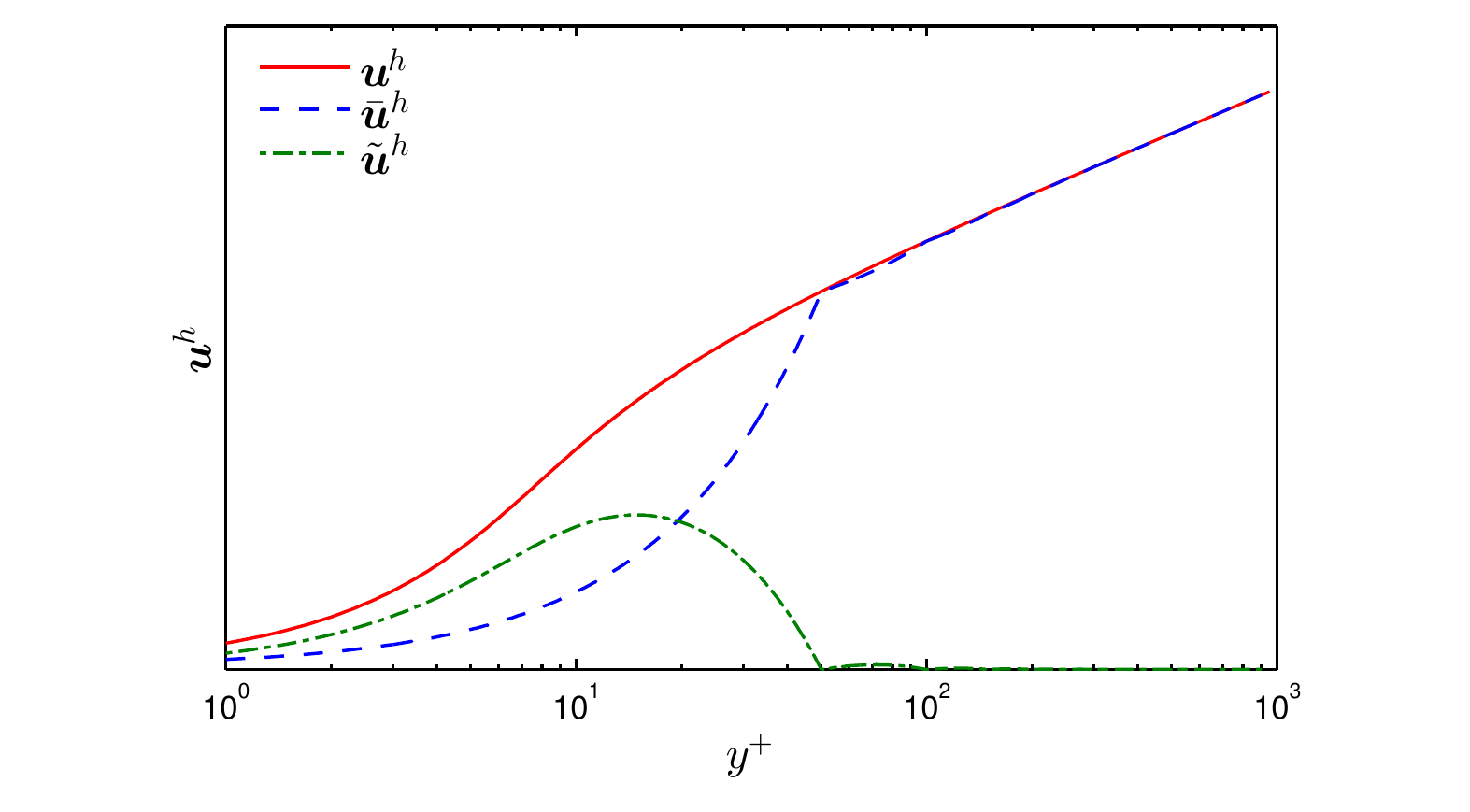}
\caption{Decomposition of the mean velocity in a buffer layer into the linear and enrichment components.}
\label{fig:decomp}
\end{figure}

An appropriate choice of the enrichment function $\psi(\bm{x},t)$ is the key feature of the overall methodology. This function provides the opportunity to include information a priori known about boundary layers in the function space without prescribing the solution itself. We propose to enrich the function space with an empirical single analytic function for the law-of-the-wall including viscous sublayer and inner layer. This idea follows the paradigm of DES stating that not all turbulent scales need to be resolved at the wall but only their ensemble-averaged solution is computed. In contrast to standard methods for DES, the resolution in wall-normal direction may be very coarse if the function space is capable of resolving the mean gradient and the resolution requirements are essentially independent of wall units. The decomposition of the proposed function space into a linear standard and enrichment component is visualized in figure \ref{fig:decomp}.

%spalding's law decomposed

Such mean velocity profiles have for example been suggested by Reichardt \cite{Reichardt51} or more widely known by Spalding \cite{Spalding61}. They satisfy the boundary conditions at the wall  $\bm{u}(y=0)=0$ and $\frac{\partial u^+}{\partial y^+}|_{y=0}=1$ (see e.g. Dean \cite{Dean76}) and may with the latter even predict the wall shear stress accurately.

\begin{figure}
\center
\includegraphics[scale=0.6]{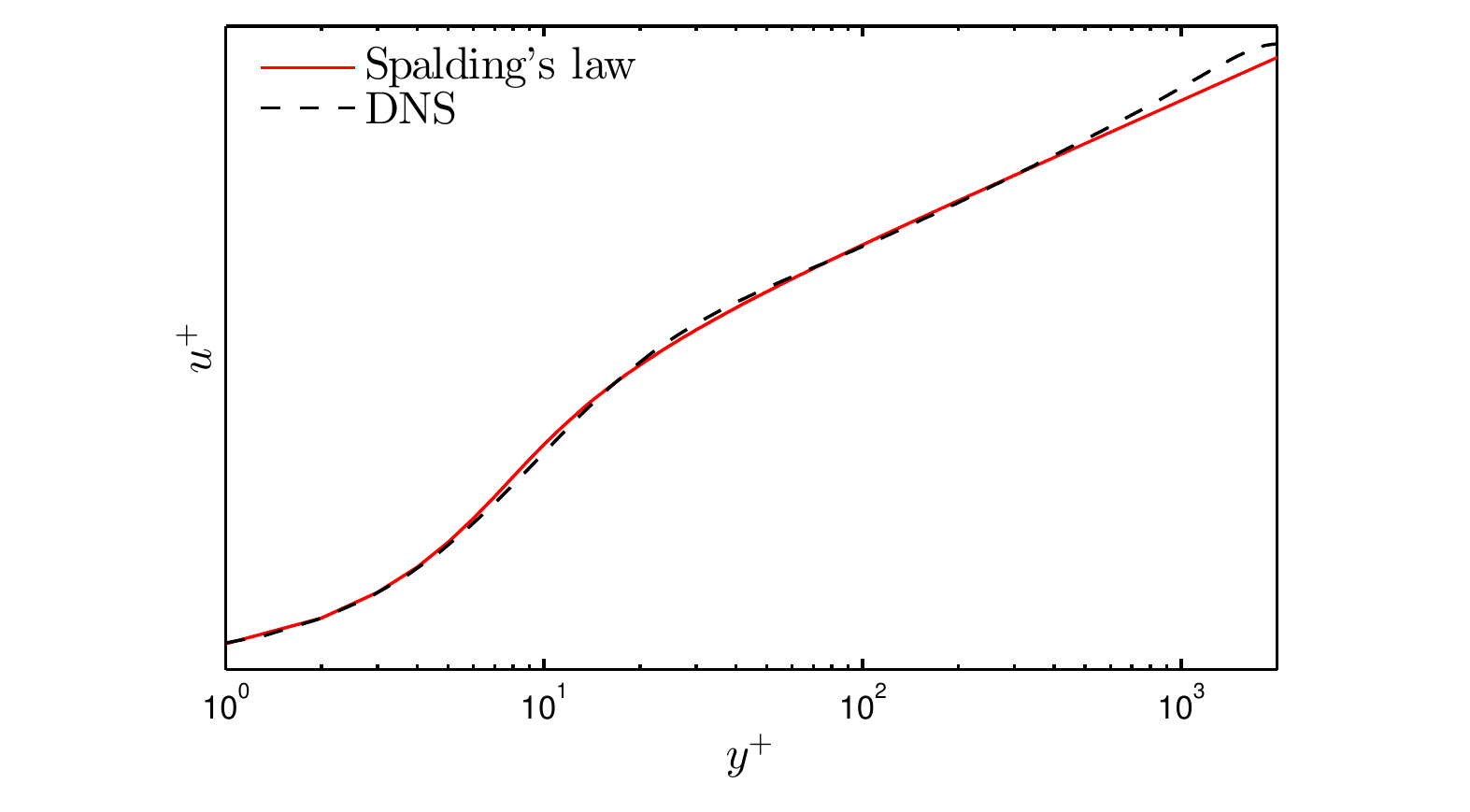}
\caption{Comparison of the law-of-the-wall with DNS data of a turbulent channel flow at $Re_{\tau}=2003$.}
\label{fig:spalddns}
\end{figure}

The enrichment function proposed in this work is a minor modification of the law-of-the-wall by Spalding 
\begin{equation}
y^+(y,\tau_w)=\frac{\psi}{\kappa}+e^{-\kappa B}(e^{\psi}-1-\psi-\frac{\psi^2}{2!}-\frac{\psi^3}{3!}-\frac{\psi^4}{4!})
\label{eq:spald}
\end{equation}
where the common formulation is recovered with $u^+=\frac{\psi}{\kappa}$. The constants $\kappa=0.41$ and $B=5.17$ by Dean \cite{Dean78} are applied. The only remaining parameters are the distance from the wall $y$ and the wall shear stress $\tau_w$ included in the definition of the wall coordinate
\begin{equation}
y^+(y,\tau_w)=\frac{y}{\nu}\sqrt{\frac{\tau_w}{\rho}}
\end{equation}
where the density is denoted $\rho$. Close agreement of Spalding's law-of-the-wall with DNS data of turbulent channel flow at $Re_{\tau}=2003$ by Hoyas and Jim{\'e}nez \cite{Hoyas06} is observed in figure \ref{fig:spalddns} in the viscous sublayer and logarithmic region. The modification of the law-of-the-wall by Dean \cite{Dean76} to also describe the wake of the channel flow is not taken into consideration here.

%spalding's law compared with DNS

In the final method a discrete version $y^{+,h}(y^h,\tau_w^h)$ is used. The discrete distance from the wall $y^h$ is defined as the distance of each node $y_{B}$, with $B \in N_{enr}^{\bm{u}}$, to the closest node at the wall in $N_{D}^{\bm{u}} \subset N_{enr}^{\bm{u}}$
\begin{equation}
y^h=\sum\limits_{B \in N_{enr}^{\bm{u}}} N_{B}^{\bm{u}}  y_{B}
\label{eq:ydiscret}
\end{equation}
forming a robust procedure even for surfaces where the wall-normal vector is not unique.

For application with the incompressible Navier-Stokes equations, expressions for the first and second derivatives of the enrichment with respect to cartesian coordinates are required. Their derivation is straight forward and included in \ref{sec:appendix}.

\subsection{Adaptivity in space and time}
\label{sec:adaptivity}
The wall shear stress being the only parameter of the function space represents an advantageous characteristic. It is known for many canonical flows in advance, such as for turbulent channel flow in a mean sense, and the pre-supposed value could be applied for computation of the wall model. However, in general flows, the tractions are not a priori known and it is desirable that the shape functions adapt to local fluctuations, regions of varying wall shear stresses and their temporal evolution. Therefore, it is suggested to explicitly compute the tractions and impose them on the wall model including spatial and temporal adaptation.

Several methods are available for determining the instantaneous shear stresses, such as the common gradient-based approach via the derivative of the wall-parallel velocity component with respect to the wall distance, given in discrete form
\begin{equation}
\tau_w^h= \sum\limits_{B \in N_D^{\bm{u}}} N_B^{\bm{u}} \nu \rho \Big\rVert\frac{\partial \bm{u}_{\|}^h}{\partial y}\bigg|_{\bm{x}_B}\Big\rVert_2.
\end{equation}

An alternative method common in the finite element method is calculating the wall shear stress via a nodal wall-parallel force vector $\bm{r}_{\|B}^{\bm{v}}$ on the Dirichlet boundary $\Gamma_D$ divided by the nodally defined local area $A_B$ and interpolated with the standard finite element expansion
\begin{equation}
\tau_w^h=\sum\limits_{B \in N_D^{\bm{u}}} N_B^{\bm{u}}  \frac{\lVert\bm{r}_{\|B}^{\bm{v}}\rVert_2}{A_B}
\label{eq:fbased}
\end{equation}
with the norm $\lVert\cdot\rVert_2$ of the three components corresponding to the space dimensions.
The force vector equals the right-hand-side residual vector of the final matrix system and is discussed in Section \ref{sec:matrixform}. The nodal area is given as the integral of the standard partition of unity on the boundary
\begin{equation}
A_B=\int_{\Gamma_D} N_B^{\bm{u}} dA.
\end{equation}

Both of these methods represent an accurate definition of the momentary traction and are exactly equivalent for the continuous case but differences arise on discrete level. One of the differences is that the latter force-based method (\ref{eq:fbased}) requires the residual to be converged to give an accurate prediction. Yet it is considered to be better consistent in the framework of FEM and chosen in this work as the standard method. The gradient-based method is applied for the first five time steps of the transient simulation  as a converged residual is not available in the first time step and the gradient-based method is more robust if the initial field is not divergence-free.

Another aspect that has to be considered in this context is that Spalding's law is a relation for mean quantities, i.e. the mean velocity is related to the average wall shear stress. The difference between applying statistical and instantaneous values of the wall shear stress becomes apparent in the force-based method: The magnitude is computed for each node in equation (\ref{eq:fbased}) resulting in a statistical over-prediction of the traction since neighboring force vectors usually are non-parallel. Therefore, it is suggested to calculate the stress via a locally averaged force field with a characteristic length scale $\alpha h$ instead of $h$ resulting in a large-scale force. Such a local averaging operation allows for spatial variations of the traction and yet local fluctuations are smoothed. This averaging is realized via level-transfer operators from plain aggregation algebraic multigrid methods for separating scales, similar to the method frequently used to explicitly separate velocity scales in LES \cite{Gravemeier09}. A discrete wall shear stress $\tau_w^{\alpha h}$ with a coarser characteristic element length $\alpha h$ as a multiple of the element length $h$ is obtained.

\begin{figure}
\centering
\begin{minipage}[b]{0.47\linewidth}
\centering
\includegraphics[scale=0.15]{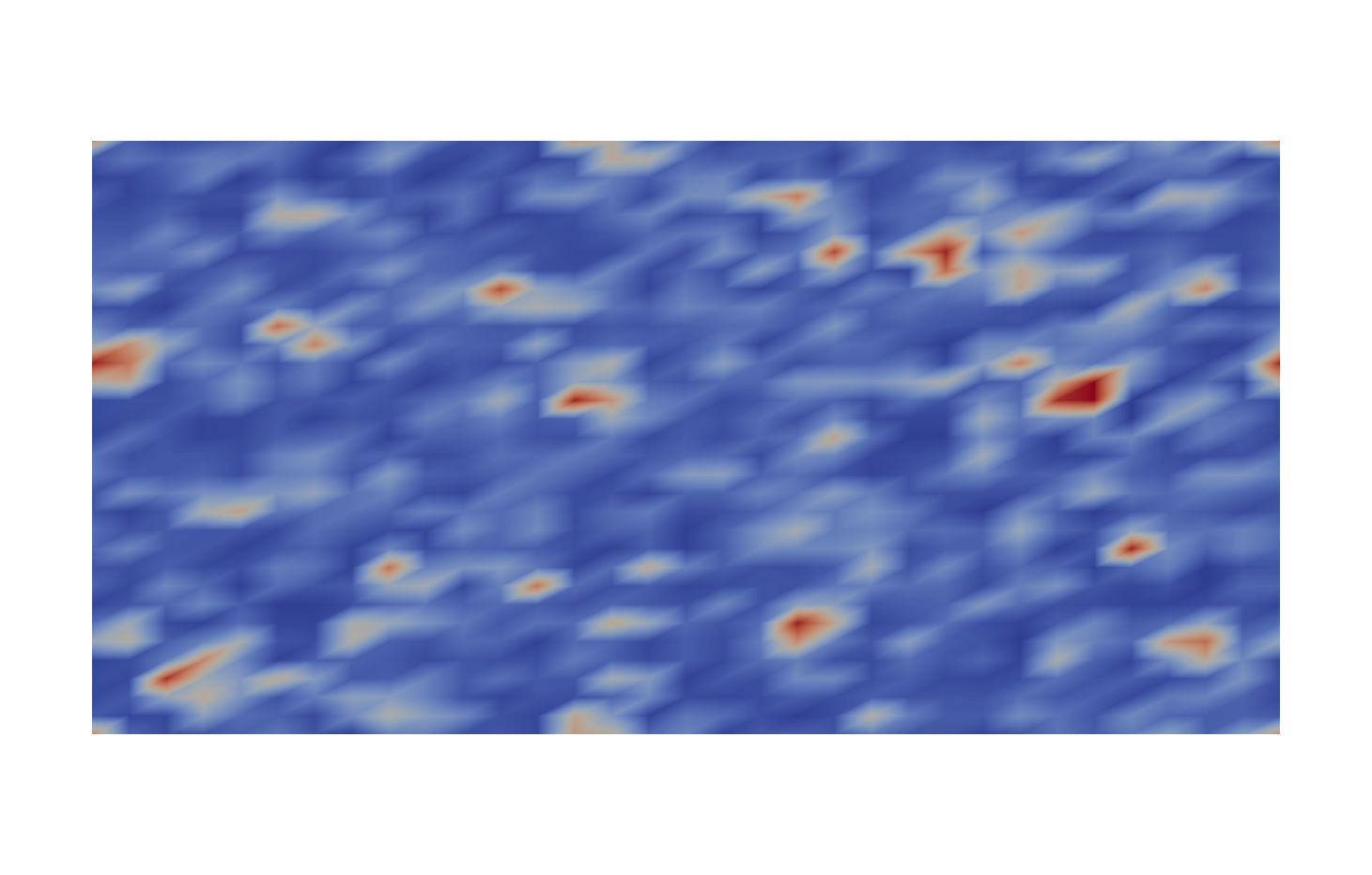}
\end{minipage}
\hspace{0.5cm}
\begin{minipage}[b]{0.47\linewidth}
\centering
\includegraphics[scale=0.15]{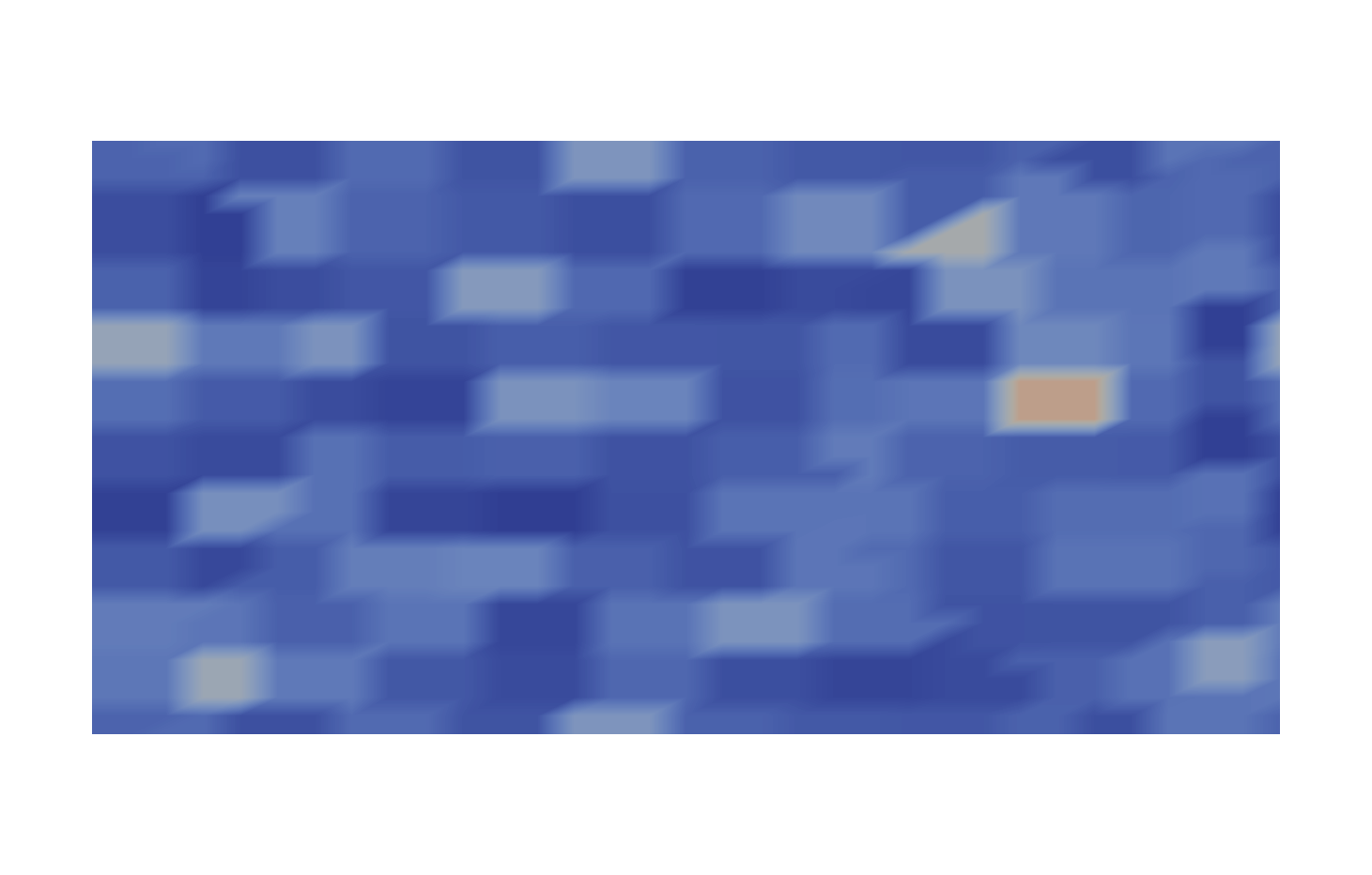}
\end{minipage}
\caption{Comparison of the wall shear stress (left) with the aggregated wall shear stress (right) of turbulent channel flow on a mesh with $32^3$ elements. Red indicates high and blue low values.}
\label{fig:wss}
\end{figure}

For this method a prolongation matrix $\bm{P}_{\alpha h}^h$ is generated and the restriction matrix is defined as the transpose of the prolongation matrix resulting in $\bm{R}_{h}^{\alpha h}=(\bm{P}_{\alpha h}^h)^T$ implying $\bm{R}_{h}^{\alpha h}\bm{P}_{\alpha h}^h=\bm{I}$ with the identity matrix $\bm{I}$. A scale-separation or aggregation operator is defined as
\begin{equation}
\bm{S}_h^{\alpha h}=\bm{P}_{\alpha h}^h \bm{R}_{h}^{\alpha h}
\end{equation}
yielding a coarse-scale force field via a vector-matrix multiplication
\begin{equation}
\bm{r}^{\bm{v},\alpha h}=\bm{S}_h^{\alpha h} \bm{r}^{\bm{v},h}.
\end{equation}
This result is applied to calculate the shear stress $\tau_w^{\alpha h}$ in equation (\ref{eq:fbased}) with a value of $\alpha=3$. Figure \ref{fig:wss} compares $\tau_w^{h}$ and $\tau_w^{3 h}$ showing that the field variable is averaged locally for the latter one but still may take larger variations into account. Thus, $\tau_w^{3h}$ is an appropriate representation of the wall-shear stress for spatial adaptation of Spalding's law.

The traction computed at the no-slip nodes is communicated to the respective off-wall nodes in $N_{enr}^{\bm{u}}$ where the node pairs are again determined via the shortest distance.

The temporal evolution of the turbulent flow results in continuous adaptation of the function space. For simplicity, the space is kept constant during the non-linear Newton iterations implying a quasi-static treatment. Updating the wall shear stress yielding new shape functions requires a subsequent L2-projection of the solution of the previous time step $n$ onto the new space of the current time step $n+1$ given in weak form as
\begin{equation}
(\tilde{\bm{v}}^{h,n+1},\tilde{\bm{u}}^{h,n+1})=(\tilde{\bm{v}}^{h,n+1},\tilde{\bm{u}}^{h,n}).
\end{equation}
Complementary vectors required by the discrete time integration procedure such as a potential acceleration vector are projected with the same matrix system.

\section{Subgrid-scale modeling}
\label{sec:subgrmodeling}
The major goal of all wall models is to utilize very coarse meshes in the near-wall region implying that a large part of the physics is not resolved but modeled. This defines a need for accurate turbulence models ensuring high-quality results. In contrast to the common approach in LES of using a filter in order to derive the unresolved part \cite{Leonard74}, a scale separation as suggested originally for large-eddy simulation by Hughes et al. \cite{Hughes00} is used here. The scale separation gives rise to unresolved scales that we model by a structural reconstruction via a multifractal subgrid scales turbulence model embedded in a residual-based variational multiscale method as outlined in the following.

\subsection{Scale separation for large-eddy simulation}
In addition to the decomposition of the solution space into standard and enriched components, the velocity space is further separated into resolved $\bm{u}^h$ and unresolved scales $\hat{\bm{u}}$ reading
\begin{equation}
\bm{u}= \bm{u}^h+\hat{\bm{u}} = \bar{\bm{u}}^h+\tilde{\bm{u}}^h+\hat{\bm{u}}.
\label{eq:threescale}
\end{equation}
The resolved scales are again identified by a characteristic element length $h$ and the unresolved scales with $\hat{(\cdot)}$.
The equivalent separation of scales of the pressure into a resolved $p^h$ as well as unresolved component $\hat{p}$ is also performed resulting in
\begin{equation}
p= p^h+\hat{p}.
\end{equation}
Direct sum decomposition of the underlying resolved and unresolved spaces is assumed as ${\mathcal{S}}_{\bm{u}}={\mathcal{S}}_{\bm{u}}^h\times\hat{\mathcal{S}}_{\bm{u}}=\bar{\mathcal{S}}_{\bm{u}}^h\times\tilde{\mathcal{S}}_{\bm{u}}^h\times\hat{\mathcal{S}}_{\bm{u}}$ and $\mathcal{S}_{p}={\mathcal{S}}_{p}^h\times\hat{\mathcal{S}}_{p}$.
Inserting these definitions into the weighted residual formulation (\ref{eq:weightedresidual}) gives rise to the following terms:
\begin{equation}
\mathcal{B}_{NS}(\bm{v},q;\bm{u}^h,p^h)+\mathcal{B}_{NS}^{lin}(\bm{v},q;\hat{\bm{u}},\hat{p})+\mathcal{C}(\bm{v};\bm{u}^h,\hat{\bm{u}})+\mathcal{R}(\bm{v};\hat{\bm{u}})=l(\bm{v})
\label{eq:vmfa}
\end{equation}
The term $\mathcal{B}_{NS}(\bm{v},q;\bm{u}^h,p^h)$ constitutes the part of the relation that is represented by the resolved solution space. The contribution $\mathcal{B}_{NS}^{lin}(\bm{v},q;\hat{\bm{u}},\hat{p})$ summarizes the linear terms dependent on the subgrid scales $\hat{\bm{u}}$ and $\hat{p}$:
\begin{equation}
\mathcal{B}_{NS}^{lin}(\bm{v},q;\hat{\bm{u}},\hat{p})=(\bm{v},\frac{\partial \hat{\bm{u}}}{\partial t}) - (\nabla \cdot \bm{v}, \hat{p}) + (\bm{\epsilon}(\bm{v}),2 \nu \bm{\epsilon} (\hat{\bm{u}}))+(q,\nabla \cdot \hat{\bm{u}})
\label{eq:linterm}
\end{equation}
The non-linear convective term gives rise to the cross- and Reynolds stresses as 
\begin{equation}
\mathcal{C}(\bm{v};\bm{u}^h,\hat{\bm{u}})=(\bm{v},\bm{u}^h \cdot \nabla \hat{\bm{u}}+\hat{\bm{u}} \cdot \nabla \bm{u}^h)
\label{eq:cross}
\end{equation}
and
\begin{equation}
\mathcal{R}(\bm{v};\hat{\bm{u}})=(\bm{v},\hat{\bm{u}} \cdot \nabla \hat{\bm{u}}).
\label{eq:reynolds}
\end{equation}

A basic characteristic of the variational multiscale method is that the solution and weighting function spaces have the same structure, i.e. the spaces for the weighting function may likewise be decomposed into the corresponding resolved and unresolved contributions:
\begin{equation}
\bm{v}= \bm{v}^h+\hat{\bm{v}} = \bar{\bm{v}}^h+\tilde{\bm{v}}^h+\hat{\bm{v}}
\end{equation}
\begin{equation}
q= q^h+\hat{q}
\end{equation}
Since equation (\ref{eq:vmfa}) is linear with respect to the weighting functions, it may be separated and as usual only the resolved scale component is taken into further consideration:
\begin{equation}
\mathcal{B}_{NS}(\bm{v}^h,q^h;\bm{u}^h,p^h)+\mathcal{B}_{NS}^{lin}(\bm{v}^h,q^h;\hat{\bm{u}},\hat{p})+\mathcal{C}(\bm{v}^h;\bm{u}^h,\hat{\bm{u}})+\mathcal{R}(\bm{v}^h;\hat{\bm{u}})=l(\bm{v}^h)
\label{eq:discretefinal}
\end{equation}

This result still contains the unresolved scale quantities $\hat{\bm{u}}$ and $\hat{p}$ which are unknown and have to be modeled. In the following sections, this is done via multifractal scale similarity as well as residual-based modeling.

\textbf{Remark:} The enrichment approach presented in this paper may also be interpreted as a separation of the solution vector in three scale groups as for example described by Gravemeier et al. \cite{Gravemeier04} and indicated in relation (\ref{eq:threescale}). The three scales are represented by the standard resolved scale $\bar{\bm{u}}^h$, the enrichment scale $\tilde{\bm{u}}^h$ as well as the unresolved scale $\hat{\bm{u}}$. With regard to LES, the standard scale resolves large eddies that are at least of the size of the characteristic element length. The enrichment scale, however, represents flow features in a statistical sense and without resolving large eddies in the near-wall region explicitly. The physical interpretation of the unresolved scales are fluctuations on subgrid level. This interpretation and the framework of the variational multiscale method would allow different turbulence models tailored for the respective requirements of the scale. The turbulence models presented in the following show to be sufficiently general and adapt to the specific requirements in different regions of the domain such that a three-scale turbulence model is not necessary.

\subsection{Subgrid modeling with multifractal subgrid scales}
The cross and Reynolds stress terms (\ref{eq:cross}) and (\ref{eq:reynolds}) are modeled explicitly by reconstruction of the unresolved scale $\hat{\bm{u}}$ via a multifractal subgrid-scale model as proposed by Rasthofer and Gravemeier \cite{Rasthofer12}. The multifractal subgrid scale model follows the idea that turbulence originates from repeated stretching and folding of vortical structures and that this process is scale-invariant. The model attempts to reconstruct the subgrid-scale vorticity and computes the subgrid velocity through the law of Biot-Savart, indicating that the large eddies of the flow have to be resolved explicitly. For a detailed derivation of the governing relations it is referred to Burton and Dahm \cite{Burton05}.

The subgrid velocity scales with the small-resolved velocity field $\delta\bar{\bm{u}}^h$ and a proportionality factor $B$ as
\begin{equation}
\hat{\bm{u}}\approx B \delta\bar{\bm{u}}^h.
\label{eq:subgrvel}
\end{equation}
The small-scale velocity is determined by an explicit filtering of the standard FE part of the resolved velocity yielding
\begin{equation}
\bm{u}=\bar{\bm{u}}^{\alpha h} + \delta\bar{\bm{u}}^h+\tilde{\bm{u}}^h+\hat{\bm{u}}
\end{equation}
for the overall composition of the velocity space. The large-scale velocity field $\bar{\bm{u}}^{\alpha h}$ is identified by a length scale of $\alpha h$ as a multiple of the element length.
This decomposition is chosen due to the physical interpretation of the standard finite element space as eddies while the enrichment space represents a statistical velocity profile that does not resolve eddies by nature.

Explicit scale separation of $\bar{\bm{u}}^{\alpha h}$ and $\delta\bar{\bm{u}}^h$ is performed via a plain aggregation algebraic multigrid method as proposed in \cite{Gravemeier09} and applies similar relations as used for smoothing of the wall shear stress in Section \ref{sec:adaptivity}. The standard parameter $\alpha=3$ is applied.

The proportionality factor in (\ref{eq:subgrvel}) is given as
\begin{equation}
B=C_{sgs}(1-\alpha^{-\frac{4}{3}})^{-\frac{1}{2}}2^{-\frac{2N}{3}}(2^{\frac{4N}{3}}-1)^{\frac{1}{2}}.
\label{eq:b}
\end{equation}
In the current application of convection-dominated high-Reynolds-number flow, the constant $C_{sgs}=0.15$ is chosen. This value is significantly lower than the one suggested in \cite{Rasthofer12}, which is since the near-wall limit as suggested in the original publication is not considered here. $B$ is evaluated at the quadrature points during evaluation of the discrete formulation (\ref{eq:discretefinal}). The number of cascade steps $N$ from the smallest resolved scales of size $h$ to the viscous scale $\lambda_v$ is approximated via the local element Reynolds number $Re_h=\frac{\lVert\bm{u}^h\rVert_2h}{\nu}$ and a proportionality constant $c_v$ resulting in
\begin{equation}
N=log_2(\frac{h}{\lambda_v})=log_2(c_v Re_h^{\frac{3}{4}}).
\end{equation}
A value for the proportionality constant $c_v=0.1$ is used, which is close to the value of $\frac{1}{12.3}$ determined experimentally by Mullin and Dahm \cite{Mullin06} and $h$ is approximated by the cube root of the local element volume.

The final result for the modeled cross and Reynolds stresses (\ref{eq:cross}) and (\ref{eq:reynolds}) with the presented relation for the subgrid-scale velocity (\ref{eq:subgrvel}) is
\begin{equation}
\mathcal{C}(\bm{v}^h;\bm{u}^h,\hat{\bm{u}})\approx(\bm{v}^h,B(\bm{u}^h\cdot\nabla\delta\bar{\bm{u}}^h+\delta\bar{\bm{u}}^h\cdot \nabla \bm{u}^h))
\end{equation}
and
\begin{equation}
\mathcal{R}(\bm{v}^h;\hat{\bm{u}}) \approx (\bm{v}^h,B^2(\delta\bar{\bm{u}}^h\cdot\nabla \delta\bar{\bm{u}}^h)).
\end{equation}

\subsection{Residual-based modeling}

The multifractal subgrid scale model as presented in the previous section enables reconstruction of the subgrid velocity field and by that models the cross and Reynolds stress terms. As reported in \cite{Burton05b}, the model allows both for dissipation and backscatter of energy resulting in potentially de-stabilizing effects. Therefore, as suggested in \cite{Rasthofer12}, it is embedded in the residual-based variational multiscale method providing a stable numerical method. 

The remaining linear terms of the scale separation (\ref{eq:linterm}) are approximated with
\begin{equation}
\mathcal{B}_{NS}^{lin}(\bm{v}^h,q^h;\hat{\bm{u}},\hat{p}) \approx \underbrace{(\bm{u}^h \nabla \cdot \bm{v}^h, \tau_M \bm{R}_{M}^h)}_{\text{SUPG}}+ \underbrace{(\nabla \cdot \bm{v}^h,\tau_C \bm{R}_{C}^h)}_{\text{grad-div}}+\underbrace{(\nabla q^h,\tau_M \bm{R}^h_M)}_{\text{PSPG}}
\end{equation}
The included terms consist in a Streamline/Upwind Petrov-Galerkin (SUPG), a grad-div (grad-div) and a Pressure Stabilizing Petrov-Galerkin (PSPG) term.
The SUPG term stabilizes the method with respect to convection by introducing a certain amount of artificial dissipation \cite{Brooks82}. Better fulfillment of the divergence-free constraint (\ref{eq:conti}) and improved convergence of the iterative solver is obtained via the grad-div term \cite{Olshanskii09} which also introduces a certain amount of dissipation in the system. The PSPG contribution enables circumventing the inf-sup condition (see e.g. \cite{Fortin91}) and allows equal-order interpolation \cite{Tezduyar92}.

The momentum residual $\bm{R}_{M}^h$ is defined as 
\begin{equation}
\bm{R}_{M}^h = \frac{\partial \bm{u}^h}{\partial t} + \bm{u}^h \cdot \nabla \bm{u}^h + \nabla p^h -  2 \nu \nabla \cdot \bm{\epsilon} (\bm{u}^h) - \bm{f}^h
+B(\bm{u}^h\cdot\nabla\delta\bar{\bm{u}}^h+\delta\bar{\bm{u}}^h\cdot \nabla \bm{u}^h)+B^2(\delta\bar{\bm{u}}^h\cdot\nabla \delta\bar{\bm{u}}^h).
\label{eq:momentres}
\end{equation}
In contrast to \cite{Rasthofer12}, it is suggested to include the modeled cross- and Reynolds stress terms (\ref{eq:cross}) and (\ref{eq:reynolds}) in the residual for better consistency. The discrete continuity residual $\bm{R}_{C}^h$ is
\begin{equation}
\bm{R}_{C}^h = \nabla \cdot \bm{u}^h.
\label{eq:contires}
\end{equation}

The stabilization parameters $\tau_M$ and $\tau_C$ are designed to take into account the non-polynomial character of the element space. A definition inspired, among others, by Codina \cite{Codina02} and Gravemeier et al. \cite{Gravemeier10} is chosen including, as usual, a transient, convective and viscous contribution for $\tau_M$ as
\begin{equation}
\tau_M = \frac{1}{\frac{1}{\Delta t}+2\sqrt{\frac{\lambda^h}{3}}\lVert\bm{u}^h\rVert_2+4\lambda^h \nu}
\label{eq:momstab}
\end{equation}
with the time step $\Delta t$ and a reciprocal scaling of $\tau_M$ and $\tau_C$ yielding
\begin{equation}
\tau_C = \frac{1}{4\lambda^h\tau_M}.
\label{eq:contistab}
\end{equation} 
It is noted that $\bigg(2\sqrt{\frac{\lambda^h}{3}}\bigg)^2 \leq 4\lambda^h$ which has been reported to be a requirement for example in \cite{Codina02}.

The parameter $\lambda^h$ generally incorporates the characteristics of the element, for example the polynomial order of the underlying function space. For standard Lagrangian elements with well-defined polynomial order, such as the non-enriched elements, values are for instance for the polynomial orders $p=\{1,2,3\}$ given as $\lambda^h=\{\frac{3}{h^2},\frac{12}{h^2},\frac{60}{h^2}\}$ with the characteristic element length $h$ \cite{Harari92,Franca92}. In the current application, element spaces of the enriched elements are non-polynomial making it impossible to specify appropriate values a priori. Therefore, an element-specific value for $\lambda^h$ is determined consistently via inverse estimate as suggested by Harari and Hughes \cite{Harari92} ensuring stability for convection-dominated flows by solving the local generalized eigenvalue problem for its maximum eigenvalue $\lambda^h$ and $v^h$ given as
\begin{equation}
(\Delta {w}^h, \Delta {v}^h)_{\Omega^e}- \lambda^h(\nabla {w}^h,\nabla {v}^h)_{\Omega^e} = 0.
\label{eq:invest}
\end{equation}
$\Omega^e$ represents the element domain and the solution $v^h$ and weighting function space $w^h$ are defined similarly as the enriched velocity space (\ref{eq:space}), e.g. for $v^h$:
\begin{equation}
v^h(\bm{x},t)=\bar{v}^h(\bm{x},t)+\tilde{v}^h(\bm{x},t)
\end{equation}
The standard and enrichment parts are given with a single degree of freedom per node as
\begin{equation}
\bar{v}^h(\bm{x},t)=\sum_{B \in N^{\bm{u}}} N_B^{\bm{u}}(\bm{x}) \bar{v}_B
\end{equation}
and
\begin{equation}
\tilde{v}^h(\bm{x},t)=\sum_{{B} \in N_{enr}^{\bm{u}}} N_{{B}}^{\bm{u}}(\bm{x}) (\psi(\bm{x},t)-\psi(\bm{x}_{{B}},t))  r^h(\bm{x}) \tilde{v}_{{B}}. 
\end{equation}

Evaluating the inverse estimate with only one degree of freedom per node results in matrix dimensions of only $16 \times 16$ for the elements presently considered, such that the eigenvalue-related computation time is negligible.
A great characteristic of the presented stabilization parameter is highlighted: $\tau_M$ and $\tau_C$ are completely free of the element length if $\lambda^h$ is determined via (\ref{eq:invest}). Especially for anisotropic elements, the definition of $h$ is not obvious and many definitions have been proposed. The advantages of such a definition are also discussed for example by Franca and Madureira \cite{Franca93}.

Solely for linear elements, the standard value of $\lambda^h = \frac{3}{h^2}$ is applied with the volume-equivalent diameter $h=(\frac{6V}{\pi})^{\frac{1}{3}}/\sqrt{n_{sd}}$ with $V$ the element volume and the number of space dimensions $n_{sd}$ for simplicity \cite{Hughes86}. Due to large variations of the stabilization parameters $\tau_M$ and $\tau_C$ within one element, especially in the first element at the no-slip boundary condition, the parameters are evaluated at the quadrature points.

The present approach for turbulence modeling has been presented as a subgrid-scale model for LES assuming that the largest eddies present are resolved by the scheme. However, inside the first element at the wall, with the first off-wall node placed at $y^+>100$, this is certainly not fulfilled in the wall-region. The presented turbulence model has yet proven to be able to model the necessary subgrid scales even if the largest eddies are not resolved everywhere.

\subsection{Final discrete problem}
\label{sec:matrixform}
The final semi-discretized problem becomes
\begin{align}
&(\bm{v}^h,\frac{\partial \bm{u}^h}{\partial t}) + (\bm{v}^h,\bm{u}^h \cdot \nabla \bm{u}^h) - (\nabla \cdot \bm{v}^h, p^h) + (\bm{\epsilon}(\bm{v}^h),2 \nu \bm{\epsilon} (\bm{u}^h)) \nonumber \\
&+\underbrace{(\bm{v}^h,B(\bm{u}^h\cdot\nabla\delta\bar{\bm{u}}^h+\delta\bar{\bm{u}}^h\cdot \nabla \bm{u}^h))}_{\mathcal{C}}
+\underbrace{(\bm{v}^h,B^2(\delta\bar{\bm{u}}^h\cdot\nabla \delta\bar{\bm{u}}^h))}_{\mathcal{R}} \nonumber \\
&+\underbrace{(\bm{u}^h \nabla \cdot \bm{v}^h, \tau_M \bm{R}_{M}^h)}_{\text{SUPG}}+ \underbrace{(\nabla \cdot \bm{v}^h,\tau_C \bm{R}_{C}^h)}_{\text{grad-div}} \nonumber\\
&+(q^h,\nabla \cdot \bm{u}^h)+\underbrace{(\nabla q^h,\tau_M \bm{R}^h_M)}_{\text{PSPG}} \nonumber \\
&= (\bm{v}^h,\bm{f}^h)+(\bm{v}^h,\bm{h}^h)_{\Gamma_N}
\label{eq:semidiscret}
\end{align}
where the contributions of multifractal and residual-based subgrid-scales are labeled. The residuals $\bm{R}_M^h$ and $\bm{R}_C^h$ are defined in (\ref{eq:momentres}) and (\ref{eq:contires}), the stabilization parameters $\tau_M$ and $\tau_C$ are given in (\ref{eq:momstab}) as well as (\ref{eq:contistab}) and $B$ in (\ref{eq:b}). The terms are integrated in space applying direction-dependent Gau\ss -quadrature rules of appropriate order that enable accurate integration despite the non-polynomial function space.
Equation (\ref{eq:semidiscret}) is integrated in time utilizing a second-order accurate generalized-$\alpha$ time integration scheme including $\rho_{\infty}=0.5$ \cite{Jansen00,Gravemeier11}. Adaptive time stepping is employed such that the maximum CFL condition is kept constant at CFL=0.5 for all simulations presented.

The final matrix system is linearized and iteratively solved via a Newton-Raphson scheme yielding
\begin{equation}
\bm{K}_i^{n+1}\Delta \bm{z}_{i+1}^{n+1} = - \bm{r}_i^{n+1}
\label{eq:matrixform}
\end{equation}
for the current time step $n+1$ and non-linear iteration $i+1$, omitting the superscript $h$ for simplicity. The increment includes both velocity and pressure increments from the current non-linear iteration such that
\begin{equation}
\Delta \bm{z}_{i+1} =
\begin{bmatrix}
\Delta \bm{U}_{i+1}  \\
\Delta \bm{P}_{i+1} 
\end{bmatrix}
\begin{bmatrix}
\bm{U}_{i+1}-\bm{U}_{i}  \\
\bm{P}_{i+1}-\bm{P}_{i} 
\end{bmatrix}.
\end{equation}
The matrix $\bm{K}$ contains the linearization of all contributions of (\ref{eq:semidiscret}) except $\mathcal{C}$, $\mathcal{R}$ and the respective stabilization terms, which are treated in a fixed-point-like procedure \cite{Rasthofer12}. The residual $\bm{r}$ summarizes all terms of (\ref{eq:semidiscret}) at the previous non-linear iteration $i$. In (\ref{eq:matsplit}) $\bm{K}$ is split into four parts including $\bm{K^{vu}}$, $\bm{K}^{\bm{v}p}$, $\bm{K}^{q\bm{u}}$ and $\bm{K}^{qp}$ and $\bm{r}$ is split into two vectors $\bm{r}^{\bm{v}}$ and $\bm{r}^q$:
\begin{equation}
\bm{K} = 
\begin{bmatrix}
\bm{K^{vu}} & \bm{K}^{\bm{v}p} \\
\bm{K}^{q\bm{u}} & \bm{K}^{qp}
\end{bmatrix}
\hspace{1cm}
\bm{r}=
\begin{bmatrix}
\bm{r}^{\bm{v}}\\
\bm{r}^q
\end{bmatrix}
\label{eq:matsplit}
\end{equation}
$\bm{K^{vu}}$ contains the transient, convective and viscous term as well as terms of SUPG and grad-div. $\bm{K}^{\bm{v}p}$ comprises the pressure term and the respective part of SUPG. $\bm{K}^{q\bm{u}}$ and $\bm{K}^{qp}$ summarize the continuity contribution and the PSPG terms.  The nodal values of the momentum-residual vector $\bm{r}^{\bm{v}}$ on the Dirichlet boundary are equivalent to the nodal forces and are used to calculate the wall-shear stress $\tau_w^{\alpha h}$ in equation (\ref{eq:fbased}).

\section{Numerical examples}
\label{sec:numerical}
In this section, the performance of the presented approach is investigated for turbulent channel flow at various Reynolds numbers, flow over periodic hills and backward-facing-step flow. The latter two examples discuss the performance under separated boundary layer conditions and adverse pressure gradients.

\begin{table}
\caption{Channel flow cases and resolutions.}
\label{tab:ch_flows} 
\begin{tabular*}{\textwidth}{l @{\extracolsep{\fill}} l l l l}
\hline
Case     & $N_{x1} \times N_{x2} \times N_{x3}$  & $Re_{\tau}$    & $y_1^+$ & $N_{wm}$ \\ \hline
$Ch8wm2$   & $8 \times 8 \times 8$    & $590$; $950$; $2,000$  & $147.5$; $237.5$; $500$ & 2\\
$Ch12wm2$  & $12 \times 12 \times 12$ & $590$; $950$; $2,000$  & $98.3$; $158.3$; $333.3$  & 2 \\
$Ch16wm3$  & $16 \times 16 \times 16$ & $590$; $950$           & $73.8$; $118.8$ & 3\\
$Ch16wm2$  & $16 \times 16 \times 16$ & $2,000$                & $250$   & 2 \\
$Ch24wm3$  & $24 \times 24 \times 24$ & $950$                  & $79.2$  & 3 \\
$Ch24wm2$  & $24 \times 24 \times 24$ & $5,000$                &$416.7$  & 2 \\
$Ch32wm2$  & $32 \times 32 \times 32$ & $5,000$  & $312.5$               & 2   \\
$Ch96$    & $96 \times 96 \times 96$ & $950$    & $1.4$              & -  \\ \hline
\end{tabular*}
\end{table}

\begin{figure}
\centering
\includegraphics[trim= 5mm 0mm 5mm 0mm,clip,scale=0.60]{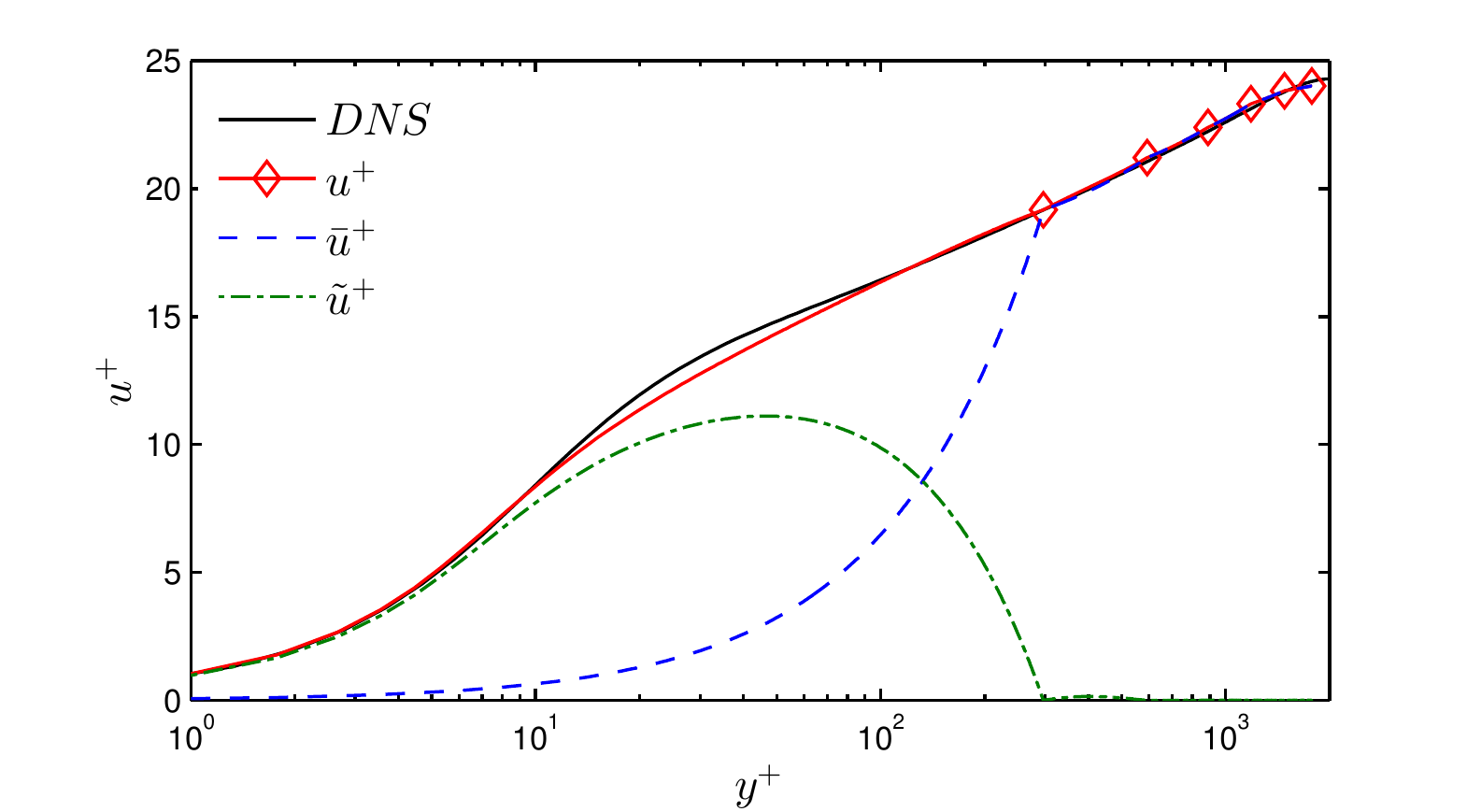}
\caption{Decomposition of the mean velocity $u^+=u_1/u_{\tau}$ with $u_{\tau}=\sqrt{\tau_w/\rho}$ of case $Ch12wm2$ at $Re_{\tau}=2,000$ into linear and enrichment part and comparison to DNS data. Symbols indicate nodes and the first off-wall node is located at $y_1^+=333.3$.}
\label{fig:ch_decomp}
\end{figure}

\subsection{Turbulent channel flow at moderate and moderately large Reynolds numbers}
A channel of the dimensions $2 \pi \delta \times 2 \delta \times \pi \delta$ in streamwise, wall-normal and spanwise direction, respectively, with periodic boundary conditions and channel-half height $\delta$ is considered. We discuss flows of friction Reynolds numbers $Re_{\tau}=590$, $950$, $2,000$ and $5,000$ on very coarse uniform meshes with $8 \times 8 \times 8$ up to $32 \times 32 \times 32$ elements, see table \ref{tab:ch_flows} for an overview. To the best of the authors' knowledge, turbulent channel flow of friction Reynolds numbers higher than $Re_{\tau}=950$ has so far never been published employing a residual-based turbulence modeling approach. The column $N_{wm}$ indicates the number of layers of enriched elements employed at the solid boundaries. Table \ref{tab:ch_flows} also compares the location of the first off-wall nodes in wall units $y_1^+$, which are located between $y_1^+ = 73.8$ and $y_1^+=500$ for the different flows and given discretizations. For comparison, we also include a simulation with resolved near-wall region without wall model at $Re_{\tau}=950$ on a discretization with $96 \times 96 \times 96$ elements. The results presented in the following are labeled according to table \ref{tab:ch_flows}. They are compared to direct numerical simulation (DNS) data of Moser, Kim and Mansour \cite{Moser99} for $Re_{\tau}=590$, Del {\'A}lamo and Jim{\'e}nez \cite{Alamo03} for $Re_{\tau}=950$ and Hoyas and Jim{\'e}nez \cite{Hoyas06} for $Re_{\tau}=2,000$. The results for $Re_{\tau}=5,000$ are compared to $u^+=\frac{1}{\kappa}ln(y^+)+B$ with $\kappa$ and $B$ defined as in Section \ref{sec:lotw}.

We commence the discussion of the results with figure \ref{fig:ch_decomp} showing the decomposition of the mean velocity of case $Ch12wm2$ at $Re_{\tau}=2,000$ similar to figure \ref{fig:decomp}. The normalized mean velocity profile $u^+=\frac{u_1}{u_{\tau}}$ with $u_{\tau}=\sqrt{\frac{\tau_w}{\rho}}$ follows DNS data closely and provides an excellent match despite the extremely coarse resolution. With the first off-wall node located at $y_1^+=333.3$, a large part of $u^+$ is in the first element represented by the enrichment part of the flow $\tilde{u}^+$ which also constitutes the largest part of the gradient at the boundary. Away from the wall in the second element layer, the contribution of the standard space $\bar{u}^+$ constitutes almost the whole solution. 

\begin{figure}
\centering
\setlength{\unitlength}{1mm}
\begin{picture}(96,82)
\put(0,0){\includegraphics[trim= 0mm 0mm 0mm 0mm,clip,scale=0.60]{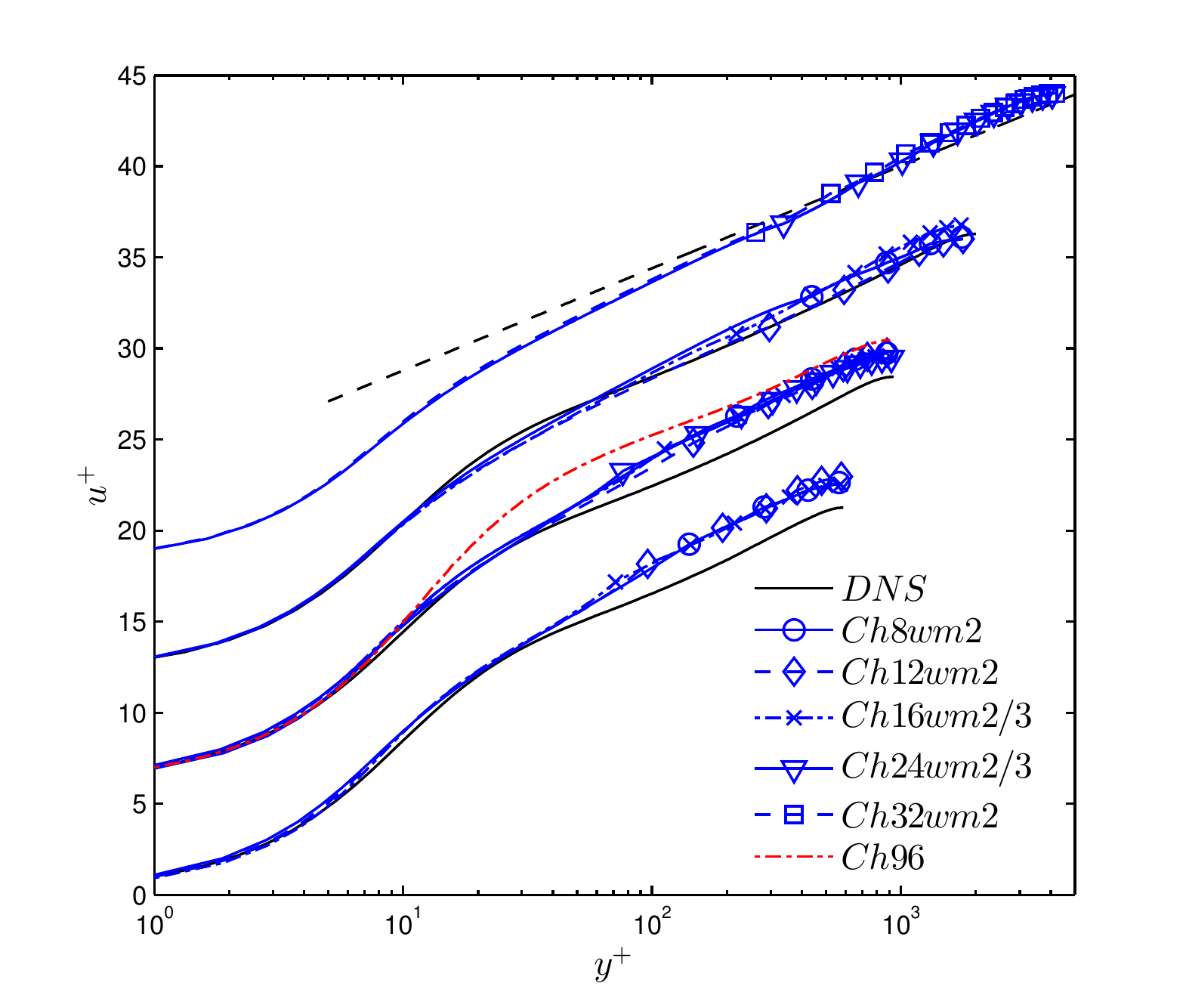}}
\put(14,15){\scalebox{.9}[1.0]{\footnotesize $Re_{\tau}$}\scalebox{.8}[1.0]{\scriptsize $=$}\scalebox{.9}[1.0]{\scriptsize $590$}}
\put(14,24){\scalebox{.9}[1.0]{\footnotesize $Re_{\tau}$}\scalebox{.8}[1.0]{\scriptsize $=$}\scalebox{.9}[1.0]{\scriptsize $950$}}
\put(14,33){\scalebox{.9}[1.0]{\footnotesize $Re_{\tau}$}\scalebox{.8}[1.0]{\scriptsize $=$}\scalebox{.9}[1.0]{\scriptsize $2,000$}}
\put(14,42){\scalebox{.9}[1.0]{\footnotesize $Re_{\tau}$}\scalebox{.8}[1.0]{\scriptsize $=$}\scalebox{.9}[1.0]{\scriptsize $5,000$}}
\end{picture}
\caption{Normalized mean velocity for $Re_{\tau}=590$, $950$, $2,000$ and $5,000$, each shifted upward by 6 units for clarity. Symbols indicate nodes.}
\label{fig:ch_umean}
\end{figure}
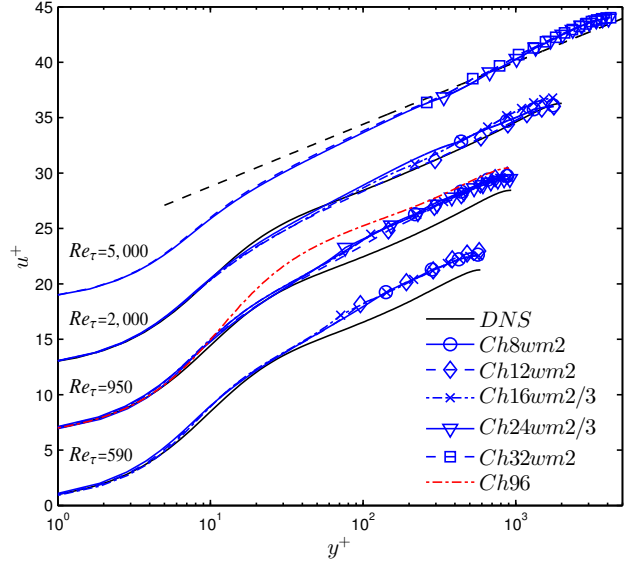

The non-dimensional mean velocity for all 13 simulations included in table \ref{tab:ch_flows} is shown in figure \ref{fig:ch_umean}. A striking independence of the mesh applied is observed for all Reynolds numbers. Even for discretizations consisting of only $8 \times 8 \times 8$ elements, where hardly the largest eddies are resolved, the results are in acceptable to very good agreement with DNS data. Further, it is noticed that the normalized mean velocity is slightly over estimated for friction Reynolds numbers $Re_{\tau}=590$ and $950$. The simulation results at $Re_{\tau}=2,000$ and $5,000$ exhibit an excellent match with reference data, however. Comparing the wall-resolved LES of case $Ch96$ at $Re_{\tau}=950$ a mean velocity of slightly lower quality than the wall-modeled LES is obtained. Very good agreement with LES data is first obtained with a mesh of $128 \times 128 \times 128$ elements as shown by Rasthofer and Gravemeier \cite{Rasthofer12}.

\begin{figure}
\centering
\includegraphics[trim= 15mm 0mm 15mm 0mm,clip,scale=0.60]{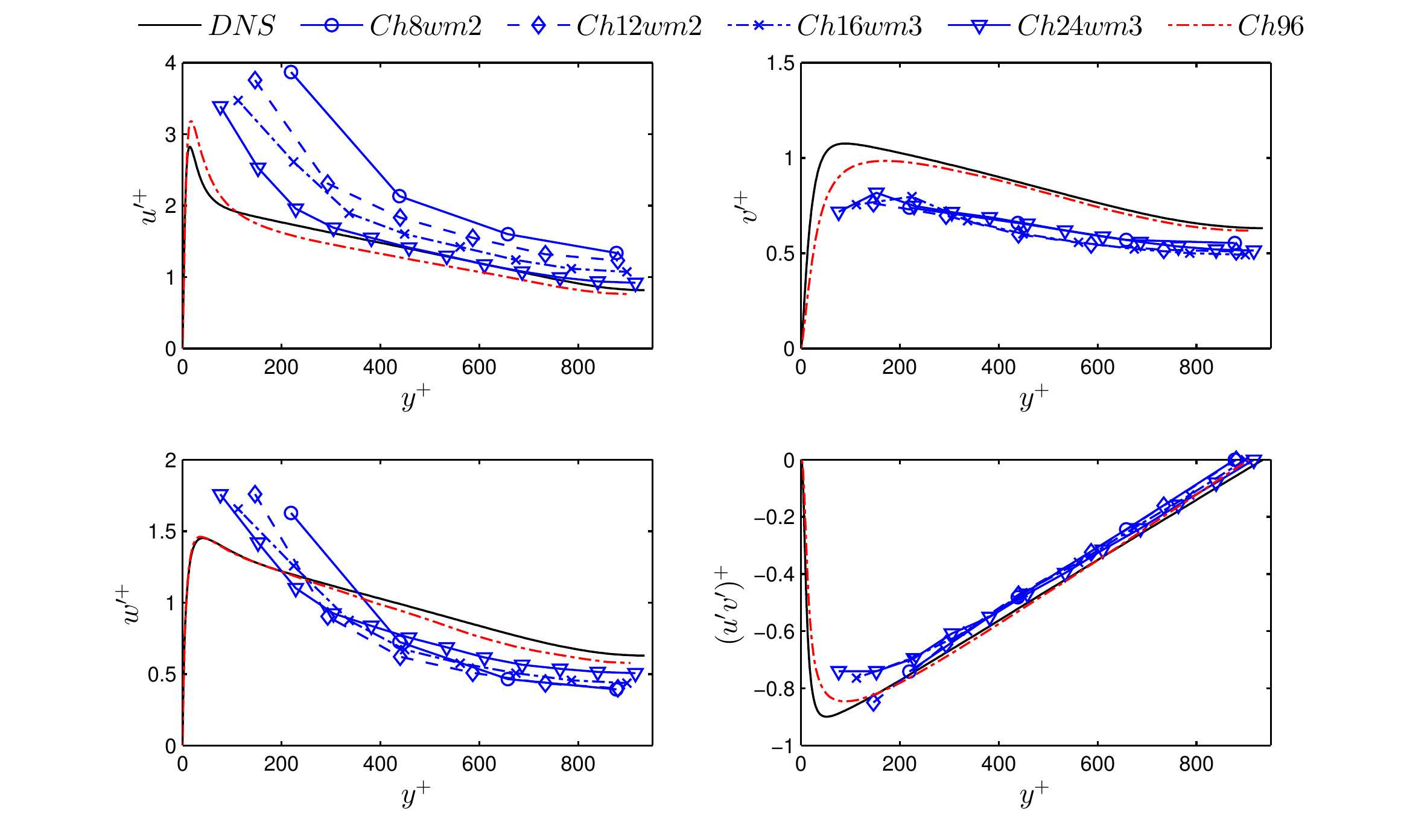}
\caption{RMS of velocity fluctuations $u^{\prime +}=rms(u_1^{\prime})/u_{\tau}$, $v^{\prime +}=rms(u_2^{\prime})/u_{\tau}$ and $w^{\prime +}=rms(u_3^{\prime})/u_{\tau}$ as well as Reynolds shear stresses $(u^{\prime}v^{\prime})^{+}=(u_1^{\prime}u_2^{\prime})/u_{\tau}^2$ for $Re_{\tau}=950$.}
\label{fig:ch_urms}
\end{figure}

The performance of the present wall model is further assed via root-mean-square (RMS) data of the fluctuations $u^{\prime +}$, $v^{\prime +}$ in wall-normal as well as $w^{\prime +}$ in spanwise direction and Reynolds shear stresses $(u'v')^+$ at $Re_{\tau}=950$ displayed in figure \ref{fig:ch_urms}. Considering $u^{\prime +}$, a distinct tendency to convergence for an increasing number of elements is observed. For $w^{\prime +}$, the predictions show a similar behavior as observed for $u^{\prime +}$ while $v^{\prime +}$ is generally predicted too small. The Reynolds shear stresses are predicted quite accurately for all discretizations. That near-wall fluctuations are not of the same quality as mean velocities is presumably inherent in the wall-modeling approach as the enrichment function constitutes a mean-velocity profile and the major part of the fluctuations is not resolved due to the coarse meshes applied. As expected, RMS data of the case $Ch96$ is in favorable agreement with DNS data as a significant amount of near-wall turbulent structures is resolved.

From the results presented in this section it is concluded that extremely coarse resolutions can be used for simulation of turbulent channel flow. The first off-wall node can be located at up to $y_1^+ = 500$ and striking results are obtained even for meshes consisting of only $8 \times 8 \times 8$ where hardly the largest eddies are resolved.

\subsection{High-Reynolds-number flow over periodic constrictions}
We consider flow over a smoothly curved 2D-periodic hill as described and analyzed numerically e.g. by Fr\"{o}hlich et al. \cite{Frohlich05} and Breuer et al. \cite{Breuer07} and experimentally by Rapp \cite{Rapp09} with a Reynolds number based on the hill height $H$ of $Re_H=10,595$ as well as $19,000$ to validate our wall modeling approach. This flow configuration includes separation at the crest of the hill, a recirculation bubble formation, reattachment as well as recovery. Wall models for LES are challenged by this flow with a strong adverse pressure gradient that causes many models to produce deficient results. For example, Chen et al. \cite{ChenHickel14} have found that their wall model based on the simplified TBLE under estimates the skin friction in the recirculation region as the convective term is neglected in that model. In the current wall modeling approach, all terms of the Navier-Stokes equations are retained such that a better performance with respect to adverse pressure gradients may be expected. Temmerman et al. \cite{Temmerman03} investigated several wall functions and subgrid closures for LES and found that the location of the separation point has major impact on the reattachment location. Also, accurate prediction of the separation point of this flow are challenging employing steady RANS simulations \cite{Jakirlic14}. Hybrid RANS/LES techniques have been analyzed by Breuer et al. \cite{Breuer08} and \v{S}ari{\'c} et al. \cite{Saric07} who have shown that the RANS/LES interface should be located inside the boundary layer on the crest of the hill. Due to the construction of the present DES technique, there is no explicit interface between the statistical and the LES region such that these problems are not expected to occur.

\begin{table}
\caption{Simulation cases and resolutions of the periodic hill. $Re_H=10,595$: $PhC$ coarse mesh with wall modeling, $PhF$ refined mesh with wall modeling, $PhFNWM$ refined mesh without wall modeling, $Froehlich\_et\_al.$ highly resolved LES, $Rapp\_EXP$ experiments, $Chen\_et\_al.\_C$ and $Chen\_et\_al.\_F$ wall modeling based on simplified TBLE and immersed interface method. $Re_H=19,000$: $PhC19$ coarse mesh with wall modeling, $PhF19$ refined mesh with wall modeling, $PhFNWM19$ refined mesh without wall modeling, $Rapp\_EXP19$ experiments.}
\label{tab:ph_flows} 
\begin{tabular*}{\textwidth}{l @{\extracolsep{\fill}} l l l l l}
\hline
Case     & $N_{x1} \times N_{x2} \times N_{x3}$  & $Re_H$ & $x_{1,sep}/H$ & $x_{1,reatt}/H$ & $N_{wm}$ \\ \hline
$PhFNWM$  & $96 \times 48 \times 48$ & $10,595$ & $0.2$  & $3.68$  & - \\
$PhC$   & $64 \times 32 \times 32$  & $10,595$  & $0.25$  & $3.77$ & 4\\
$PhF$  & $96 \times 48 \times 48$ & $10,595$ & $0.25$  & $4.91$  & 4 \\
$Froehlich\_et\_al.$  & $192 \times 128 \times 186$ & $10,595$ & $0.2$ & $4.6-4.7$ & -\\ 
$Rapp\_EXP$  & - & $10,600$ &  & $4.21$ & -\\ 
$Chen\_et\_al.\_C$  & $96 \times 64 \times 32$ & $10,595$ & $0.65$ & $4.0$ & -\\
$Chen\_et\_al.\_F$  & $192 \times 72 \times 48$ & $10,595$ & $0.5$ & $4.42$ & -\\ \hline
$PhFNWM19$  & $96 \times 48 \times 48$  & $19,000$ & $0.2$  & $3.4$  & - \\
$PhC19$   & $64 \times 32 \times 32$  & $19,000$  & $0.24$  & $2.58$ & 4\\
$PhF19$  & $96 \times 48 \times 48$  & $19,000$ & $0.26$  & $3.94$  & 4 \\
$Rapp\_EXP19$  & - & $19,000$ &  & $3.94$ & -\\ 
 \hline
\end{tabular*}
\end{table}

\begin{figure}
\begin{minipage}[b]{0.485\linewidth}
\centering
\includegraphics[trim= 0mm 0mm 0mm 0mm,clip,scale=0.17]{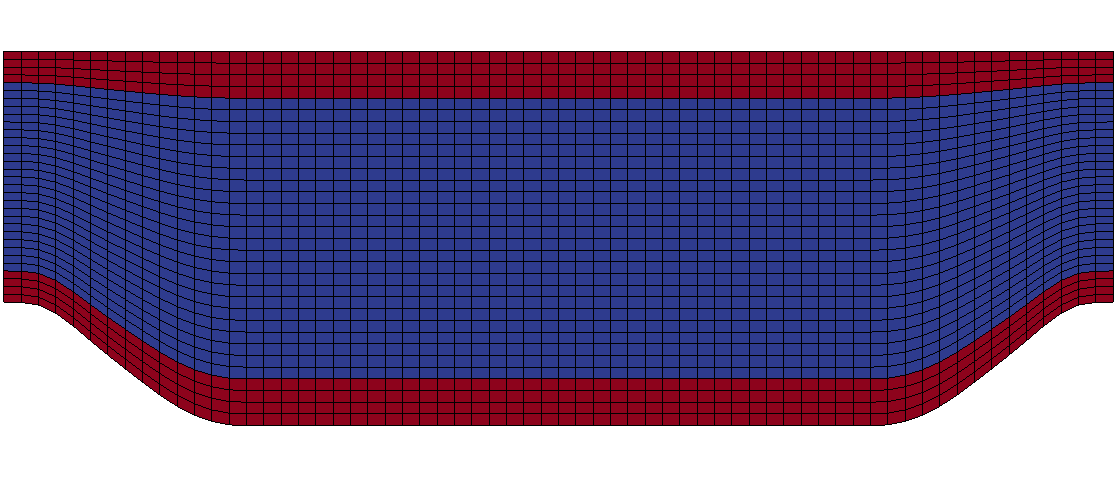}
\vspace{7mm}
\caption{Grid of case $PhC$. Enriched elements are colored red and standard linear elements blue.}
\label{fig:ph_2d}
\end{minipage}
\hspace{0.1cm}
\begin{minipage}[b]{0.485\linewidth}
\centering
\includegraphics[trim= 1mm 0mm 10mm 5mm,clip,scale=0.60]{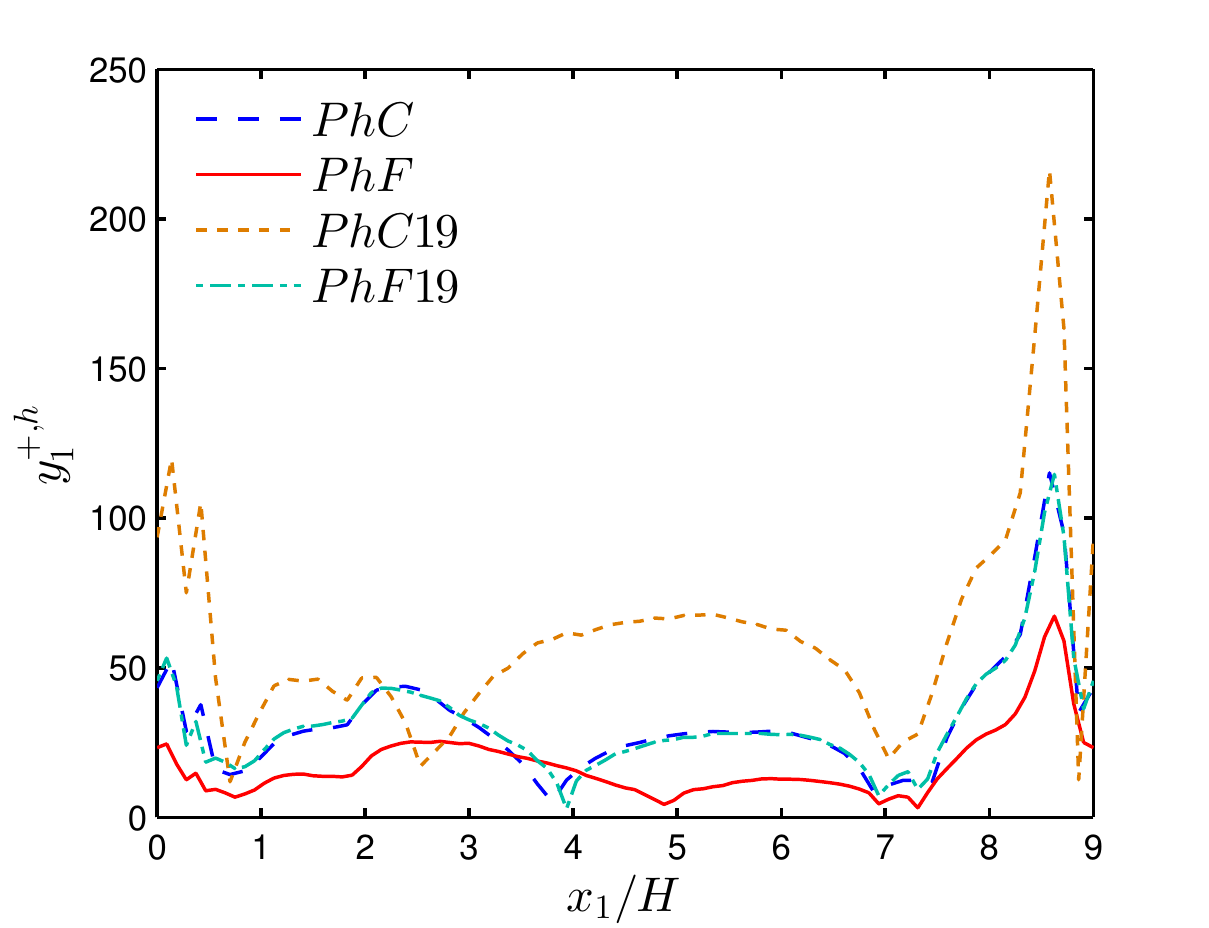}
\caption{Location of the first off-wall node in wall units over the streamwise coordinate of the periodic hill case.}
\label{fig:ph_y1}
\end{minipage}
\end{figure}

A domain of the dimensions $9H \times 3.036H \times 4.5H$ with periodic boundary conditions in the streamwise and spanwise direction and no-slip boundary conditions at top and bottom is considered. A very coarse mesh comprising $64 \times 32 \times 32$ as well as a refined, yet very coarse, mesh with $96 \times 48 \times 48$ cells with uniform grid spacings in all directions and vertical grid lines as depicted in figure \ref{fig:ph_2d} are utilized. An overview of the simulations presented is given in table \ref{tab:ph_flows} where the coarser grid is labeled $PhC$ and the finer grid $PhF$ for $Re_H=10,595$. A simulation without wall model is also investigated for comparison, which is labeled $PhFNWM$ and employs the finer mesh. Considering $Re_H=19,000$, the same meshes are applied and labeled $PhC19$, $PhF19$ and $PhFNWM19$, respectively. The number of enriched element layers is $N_{wm}=4$ for both meshes on both top and bottom wall. Figure \ref{fig:ph_y1} shows the location of the first off-wall grid point estimated as $y_1^{+,h}=\frac{y_1^h}{\nu}\sqrt{\frac{\tau_w^h}{\rho}}$ over the $x_1$-coordinate. Here, $y_1^h$ is not the actual wall distance but the distance to the closest node at the wall. The first off-wall node is located at varying distance depending on resolution and Reynolds number up to approximately $y_1^{+,h}=216$ with minima near the zero-crossings of the wall shear stress. The mass flow is kept approximately constant over the simulation time and statistics are sampled over 10,000 time steps.

An overview with respect to reference data considered is also given in table \ref{tab:ph_flows}. The results for $Re_H=10,595$ are compared to data of highly resolved LES by Fr\"{o}hlich et al. \cite{Frohlich05}, labeled as $Froehlich\_et\_al.$ as well as the coarse mesh discussed by Chen et al. \cite{ChenHickel14} ($Chen\_et\_al.\_C$) with $64$ cells in vertical direction. The separation and reattachment points are further compared to experiments by Rapp \cite{Rapp09} ($Rapp\_EXP$) and the fine mesh by Chen et al. \cite{ChenHickel14} ($Chen\_et\_al.\_F$). The results for $Re_H=19,000$ are compared to experiments by Rapp \cite{Rapp09} ($Rapp\_EXP19$). Data labeled as $Rapp\_EXP$ and $Rapp\_EXP19$ has been obtained from the ERCOFTAC QNET-CFD Wiki contributed by Rapp et al. \cite{RappERCOFTAC}.

\begin{figure}
\centering
\includegraphics[trim= 15mm 3mm 15mm 9mm,clip,scale=0.60]{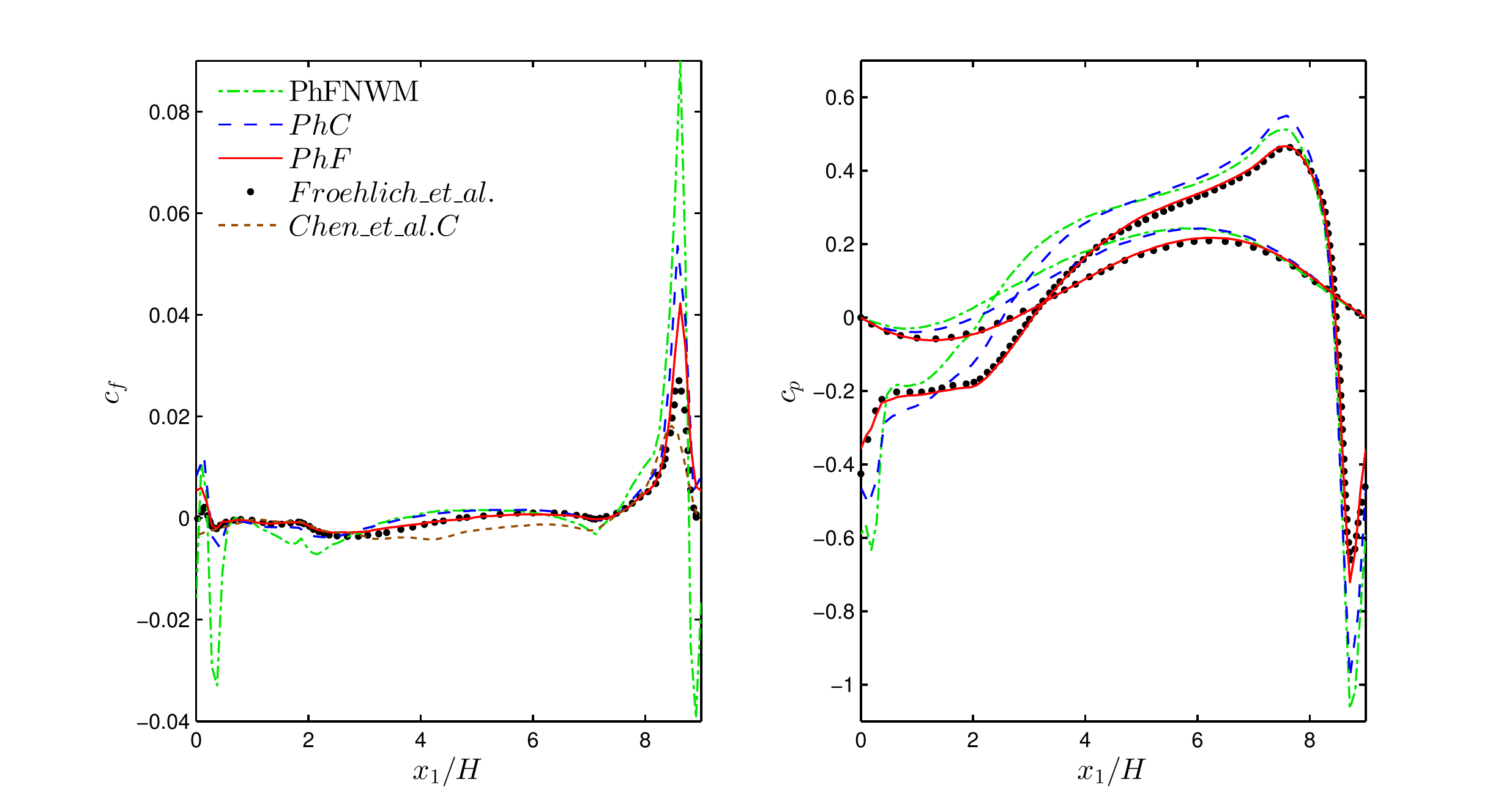}
\caption{Skin-friction (left) and pressure (right) coefficients for the flow over periodic constrictions at $Re_H=10,595$. The shallower pressure-coefficient curves correspond to the top wall.}
\label{fig:ph_cfcp}
\end{figure}

\begin{figure}
\centering
\includegraphics[trim= 15mm 30mm 15mm 22mm,clip,scale=0.60]{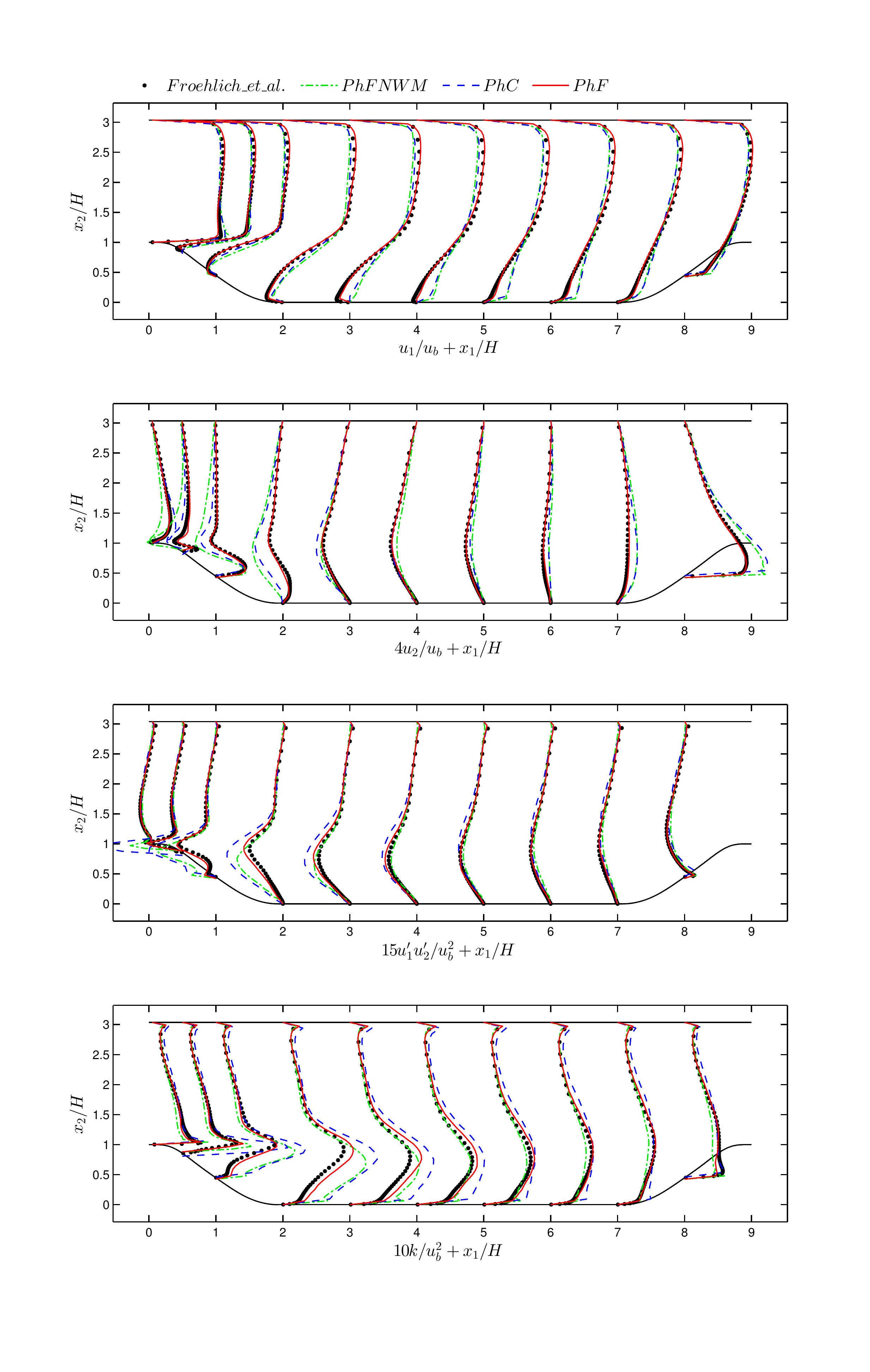}
\caption{Mean velocity $u_1$ in $x_1$ and $u_2$ in $x_2$-direction, turbulent kinetic energy $k$ as well as Reynolds shear stresses $u_1^{\prime}u_2^{\prime}$ for the periodic hill at $Re_H=10,595$.}
\label{fig:ph_uvuv}
\end{figure}

\begin{figure}
\centering
\includegraphics[trim= 15mm 20mm 15mm 13mm,clip,scale=0.60]{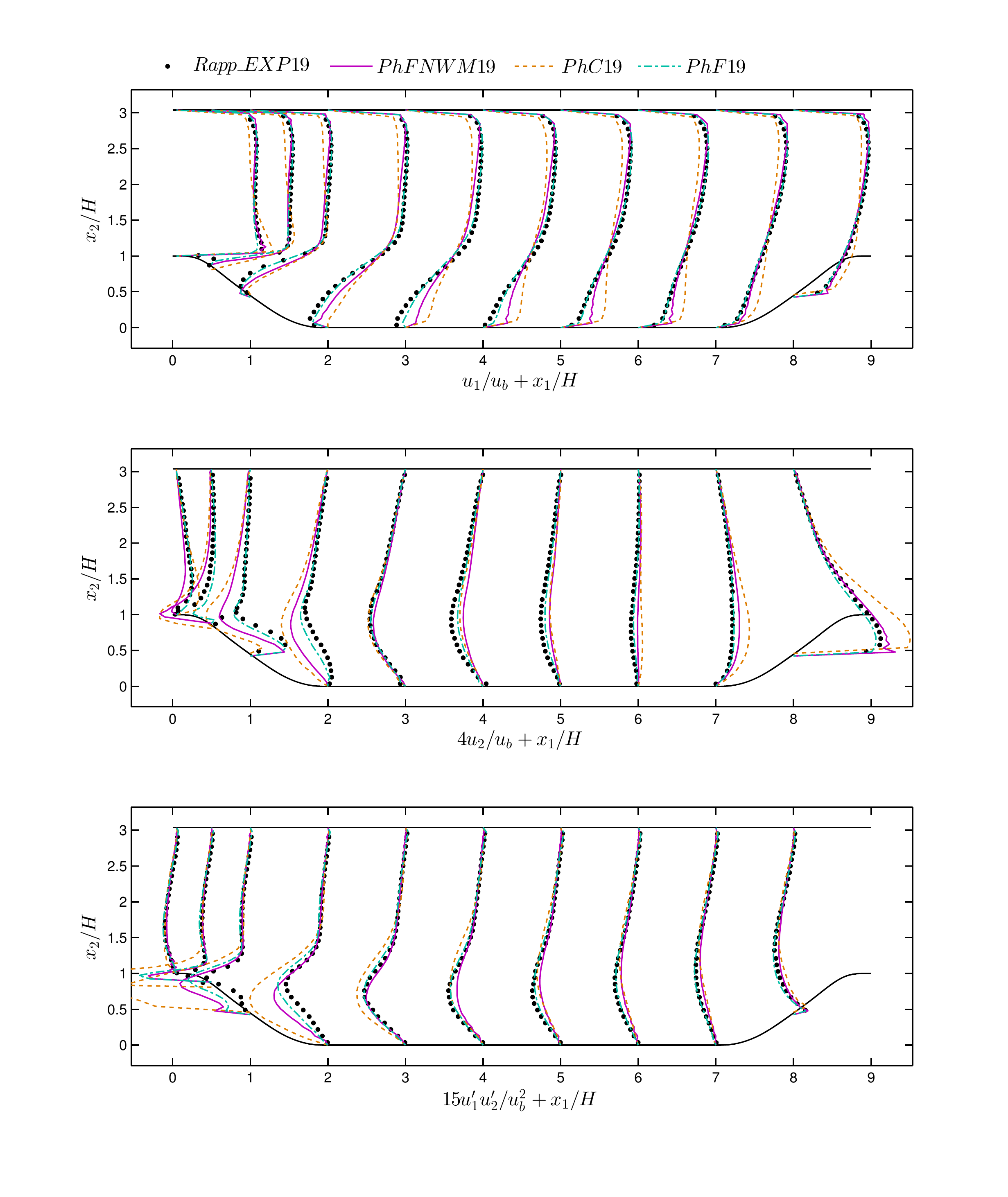}
\caption{Mean velocity $u_1$ in $x_1$ and $u_2$ in $x_2$-direction as well as Reynolds shear stresses $u_1^{\prime}u_2^{\prime}$ for the periodic hill at $Re_H=19,000$.}
\label{fig:ph_uvuv19}
\end{figure}

We start with a discussion of the results for the flow at $Re_H=10,595$. The skin-friction coefficient $c_f$ at the lower wall and pressure coefficients $c_p$ at the upper and lower wall are compared to resolved LES data in figure \ref{fig:ph_cfcp}. They are defined as
\begin{equation}
c_f=\frac{\tau_w}{\frac{1}{2} \rho u_b^2}
\label{eq:cfdef}
\end{equation}
and
\begin{equation}
c_p=\frac{p-p_{ref}}{\frac{1}{2} \rho u_b^2}
\label{eq:cpdef}
\end{equation}
with the bulk velocity $u_b$ and $\tau_w$ via the right-hand side residual (\ref{eq:fbased}). As reference pressure $p_{ref}$, the pressure at the upper wall at $x_1=0$ is chosen. The skin friction computed via wall-modeled LES is in close agreement with reference data over large parts of the domain. Solely at the crest of the hill, the peaks between $x_1=8$ and $9$ as well as $x_1=0$ and $1$ are significantly over-predicted, but improve for the case $PhF$ with higher resolution. This over-prediction might be related to the local averaging operation of the wall shear stress applied during construction of the shape functions. The minor recirculation at the top of the hill observed in highly resolved LES data is not visible in the results of $PhC$ and $PhF$ due to the coarseness of the mesh. Separation and re-attachment points are predicted accurately via the zero-crossing of $c_f$ as well and are summarized in table \ref{tab:ph_flows}. For the case $PhFNWM$ without enrichment, large discrepancies including high peaks are visible on top of the hill. In the recirculation region, the skin-friction is over estimated and the reattachment length is predicted shorter than for the cases with wall modeling. The skin-friction coefficient is also compared to the results of Chen et al. \cite{ChenHickel14}, who under estimate $c_f$ significantly due to the neglected convective term as aforementioned. 

The pressure curves of the present wall model are also in very good agreement to reference data and improve with resolution. The case $PhC$ shows minor discrepancies both at the lower and upper wall, which are due to the coarseness of the resolution. The case $PhFNWM$ over-predicts the pressure in the recirculation bubble and exhibits negative peaks on the hill crest. In the recovery region, the estimation is comparable to the coarse mesh with wall modeling $PhC$. At the upper wall, the prediction with wall model is superior compared to the one without.

Mean velocities $u_1$ in streamwise and $u_2$ in vertical direction and Reynolds-shear stresses $u_1^{\prime}u_2^{\prime}$ and the turbulent kinetic energy (TKE) $k=\frac{1}{2}(u_1^{\prime2}+u_2^{\prime2}+u_3^{\prime2})$ of the case $Re_H=10,595$ at ten stations are compared with the LES data of Fr\"ohlich et al. \cite{Frohlich05} in figure \ref{fig:ph_uvuv}. The mean velocity $u_1$ exhibits discrepancies with reference data for case $PhC$ in the reattachment and recovery region while the finer mesh $PhF$ results in a perfect match with reference data. Without wall model, $PhFNWM$ predicts $u_1$ with similar quality as the coarse mesh with wall modeling, $PhC$. Also for the mean velocity $u_2$ in vertical direction, excellent results are obtained for the fine mesh including wall modeling $PhF$. For the other cases, $u_2$ is under estimated above the recirculation bubble due to the shorter re-attachment length. The Reynolds-shear stresses $u_1^{\prime}u_2^{\prime}$ are heavily over-predicted for the cases $PhC$ and $PhFNWM$ at the crest of the hill and in the shear layer between the recirculation region and the bulk flow. Refinement leads to an excellent match with reference data for the case $PhF$. Finally, the TKE distributions are only predicted accurately everywhere with $PhF$ while $PhC$ and $PhFNWM$ over-predict its magnitude inside the recirculation bubble.

The excellent results observed for $Re_H=10,595$ motivate an application of the wall model to a higher Reynolds number. For this second assessment we choose the next larger Reynolds number for which reference data is available, which is $Re_H=19,000$, and consider the same discretizations labeled $PhC19$, $PhF19$ and $PhFNWM19$, respectively. The results in figure \ref{fig:ph_uvuv19} include mean velocities $u_1$ and $u_2$ as well as Reynolds shear stresses $u_1^{\prime}u_2^{\prime}$ and are compared to experimental data $Rapp\_EXP19$, which do not include TKE statistics. The quality of the fine mesh including wall modeling is very similar to the Reynolds number $Re_H=10,595$ discussed above and excellent throughout and also matches the reattachment length perfectly. The coarser mesh $PhC$ shows slightly worse predictions in the recirculation, reattachment and recovery region for the mean velocity $u_1$ and in the recirculation bubble for the Reynolds stresses. Also the reattachment length is predicted significantly too short which may be due to an overly coarse mesh. The case without wall model $PhFNWM19$ exhibits results of quality between the coarse and fine wall-modeled simulations. Here, another defect is highlighted: significant oscillations in the mean velocity profiles are visible especially at the lower wall and in the vicinity of the hill. Such oscillations are not visible for the wall-modeled computations and show another advantage of our wall model.

From the investigations of flow over periodic hills the conclusion may be drawn that the present enrichment-based wall model exhibits favorable characteristics with respect to separated flows as well as under adverse pressure gradients. In this flow configuration, very coarse meshes may be used, resembling the observations made for turbulent channel flow.

\subsection{Backward-facing step flow}
\begin{table}
\caption{Simulation cases and resolutions of backward-facing step flow. $BFS\_NWM$ without wall modeling, $BFS\_WM3$ with wall modeling, $BFS\_J\&D\_EXP$ experiments, $BFS\_LMK\_DNS$ DNS. }
\label{tab:bfs_flows} 
\begin{tabular*}{\textwidth}{l @{\extracolsep{\fill}} l l l l}
\hline
Case     &  $Re_h$ &  $x_{1,reatt}/h$ & $N_{wm}$ \\ \hline
$BFS\_NWM$  & $5,000$  & $13.49$  & - \\
$BFS\_WM3$   & $5,000$  & $6.78$ & 3\\
$BFS\_J\&D\_EXP$  &  $5,000$ & $6.0 \pm 0.15$  & - \\
$BFS\_LMK\_DNS$  & $5,100$ & $6.28$ & -\\ 
\hline
\end{tabular*}
\end{table}

\begin{figure}
\centering
\includegraphics[trim= 0mm 15mm 0mm 5mm,clip,scale=0.35]{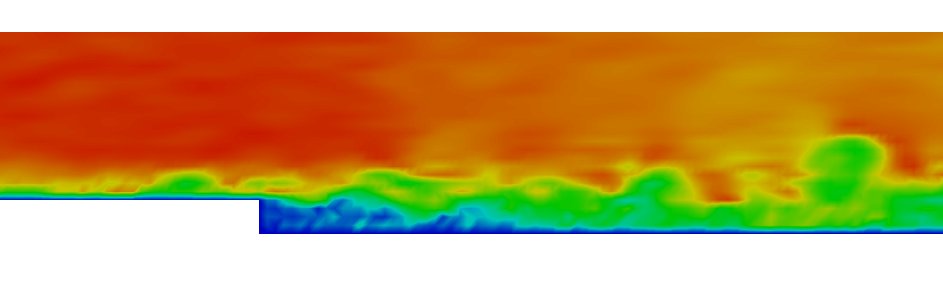}
\includegraphics[trim= 0mm 15mm 0mm 5mm,clip,scale=0.35]{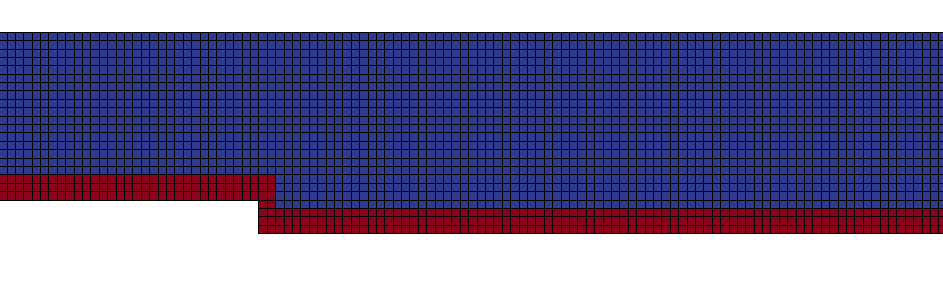}
\caption{(Top) Instantaneous velocity magnitude over the backward-facing step: red indicates high and blue low values. (Bottom) Mesh in the vicinity of the step: enriched elements are colored red and standard elements blue.}
\label{fig:bfs_vel}
\end{figure}

\begin{figure}
\centering
\includegraphics[trim= 15mm 0mm 15mm 0mm,clip,scale=0.60]{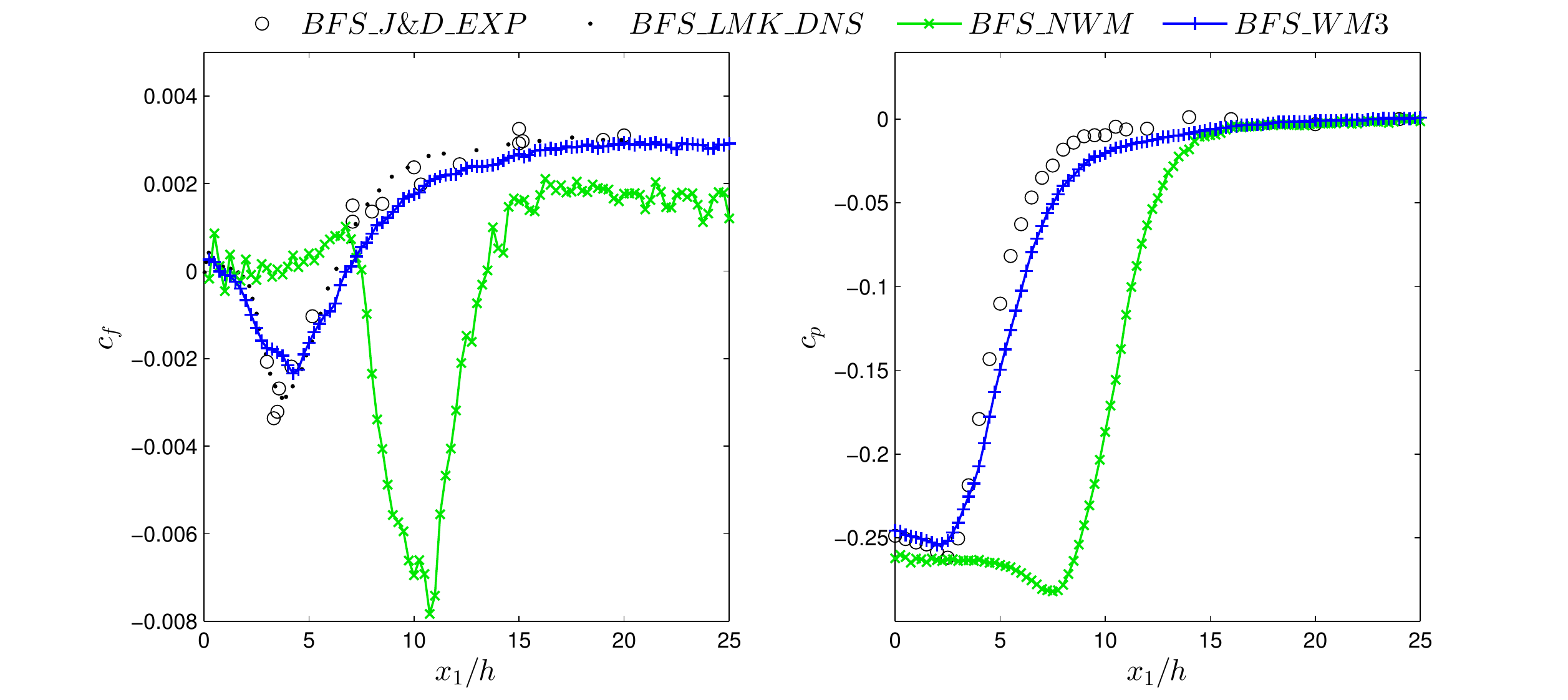}
\caption{Skin-friction (left) and pressure (right) coefficients for the flow over a backward-facing step at $Re_h=5,000$.}
\label{fig:bfs_cfcp}
\end{figure}

We assess our wall modeling approach further with flow over a backward-facing step at $Re_h=5,000$ with an expansion ratio of $ER=1.2$ as studied experimentally by Jovic and Driver \cite{Jovic94}. DNS data of a similar configuration at a Reynolds number of $Re_h=5,100$ has been presented by Le, Moin and Kim \cite{Le97} and in the context of wall modeling results have been presented e.g. by Chen et al. \cite{ChenHickel14} mentioned earlier, who encountered difficulties predicting the correct skin friction and reattachment point for this flow as well. 

The computational domain behind the step is of the dimensions $30h \times 6h \times 3h$ in streamwise, wall-normal and spanwise direction, respectively. The domain extends $30h$ upstream of the step and the velocity is prescribed at the inflow boundary using mean DNS data of a turbulent boundary layer at a similar Reynolds number \cite{Spalart88} with an additional random perturbation of $10\%$ of the center line velocity $u_c$. The inflow data is only prescribed on the standard space $\bar{\bm{u}}^h$ while the enrichment $\tilde{\bm{u}}^h$ is set to zero for simplicity. Periodic boundary conditions are applied in spanwise direction and slip boundary conditions at the top. The domain is meshed uniformly with four elements per step height $h$ in all space dimensions and for the case with wall modeling, three rows of elements at the lower wall are enriched, including the inflow region as well as the step. The resulting mesh is extremely coarse and displayed in figure \ref{fig:bfs_vel} along with a contour plot of the instantaneous velocity. For the statistical results presented in the following, the quantities are sampled for $5,000$ time steps starting after the initial transient.

An overview over the results discussed is provided in table \ref{tab:bfs_flows}. The computation including wall modeling is labeled as $BFS\_WM3$ and compared to the same mesh where the enriched elements are replaced by standard elements, labeled $BFS\_NWM$. We compare our results with the experiments by Jovic and Driver \cite{Jovic94} labeled as $BFS\_J\&D\_EXP$ and the skin friction is in addition evaluated against the DNS data by Le, Moin and Kim \cite{Le97} $BFS\_LMK\_DNS$.

\begin{figure}
\centering
\includegraphics[trim= 15mm 20mm 15mm 10mm,clip,scale=0.60]{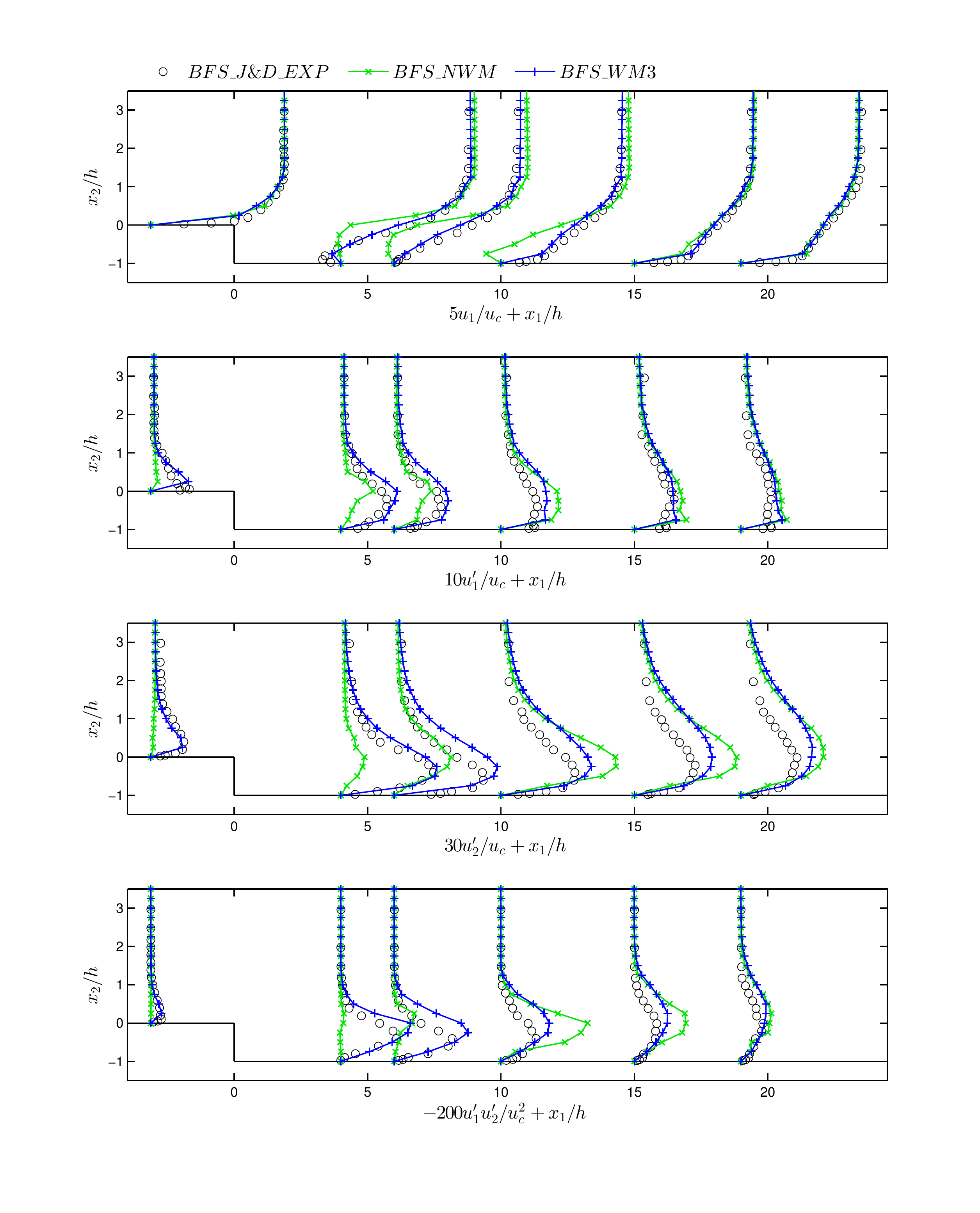}
\caption{Mean velocity $u_1$ in $x_1$ direction, RMS values of the fluctuations $u_1^{\prime}$ and $u_2^{\prime}$ in $x_2$-direction as well as Reynolds shear stresses $u_1^{\prime}u_2^{\prime}$ for the backward-facing step flow.}
\label{fig:bfs_uvuv}
\end{figure}

Again we start with a discussion of the distribution of the skin friction as well as the pressure coefficient along the lower wall. They are defined similar to (\ref{eq:cfdef}) and (\ref{eq:cpdef}) with the reference velocity $u_c$ and the reference pressure located at $x_{1}=24h$. From the results in figure \ref{fig:bfs_cfcp} it can be seen that the skin friction matches reference data very well for the case $BFS\_WM3$ with wall modeling. However, the computation without wall model $BFS\_NWM$ does not give physically reasonable results. The peak in negative skin friction is very large and shifted in streamwise direction by several step heights and, additionally, significant oscillations are observed. Accordingly, the reattachment length defined as the zero-crossing of the skin friction coefficient is predicted as $x_{1,reatt}=6.78h$ for the case with wall modeling which matches the references of $x_{1,reatt}=6.0h$ and $x_{1,reatt}=6.28h$ quite well. In contrast, the simulation without wall modeling predicts $x_{1,reatt}=13.49h$. An overview with respect to reattachment lengths for the simulations and reference data considered is also given in table \ref{tab:bfs_flows}.
The prediction of the pressure coefficient shows similar quality as for the friction coefficient. Including wall modeling, the curve follows reference data closely, while the one without wall model is delayed by several step lengths.

The mean streamwise velocity, root-mean square velocity fluctuations of the streamwise and wall-normal components and Reynolds shear stresses are displayed at six stations in figure \ref{fig:bfs_uvuv}. From these graphs it can be seen that both the velocity and fluctuations in front of the step are in good agreement with reference data implying that the simple procedure of applying inflow data gives good results. For the case without wall modeling, the fluctuations are not reproduced correctly, however, which is probably due to the extreme under-resolution.

The mean velocity behind the step matches reference data very well for the case $BFS\_WM3$ including wall modeling. It is mentioned here that the velocity is only post-processed on the element nodes which are connected with straight lines in the graph for simplicity. Therefore, the detailed velocity distribution at the second station inside the recirculation region is not shown in the graph. Without enrichment the result is barely physical and the size of the recirculation is significantly over-predicted. 

The RMS quantity $u_1^{\prime}$ is predicted very well for the case with wall modeling. Even without wall modeling the match is quite good in the recovery region but lower quantities are observed inside the recirculation. The root-mean square of the fluctuation in $x_2$-direction, $u_2^{\prime}$, is also predicted well employing the enrichment wall modeling approach. In the recovery region, small discrepancies are visible, however. We assume here that this behavior is due to the coarseness of the mesh in the shear layer above the recirculation and is not directly related to the wall model. The reference without wall model yields insufficient predictions which are faulty already at the first station.
The Reynolds shear stresses are predicted with acceptable accuracy at the first stations but an over estimation is observable around the fourth station. Without wall model, the Reynolds shear stresses are neither predicted accurately at the inflow nor behind the step.

From the backward-facing step flow investigated with and without wall model in this section we find further evidence that our wall modeling approach gives excellent results in separated flow regimes. The method is robust with respect to jumps in bounding surfaces or ambiguous wall-normal vectors. Its strengths are accurate predictions for the skin-friction and pressure coefficients as well as mean velocity profiles with very coarse meshes, but even turbulence quantities are estimated well.

\section{Conclusion}
\label{sec:conclusions}
A new approach to wall modeling for turbulent incompressible flows at moderate and high Reynolds numbers has been proposed. It is suggested to enrich the function space of the computational method with a minor modification of Spalding's law-of-the-wall such that the mean boundary layer gradient can be resolved with very coarse meshes. It is the nature of the numerical method applied, the extended finite element method, to select the most appropriate function among all functions available in its function space, in this case standard linear Lagrangian shape functions and the enrichment. Hence, Spalding's law is not prescribed but offered as alternative solution in a consistent way. The general framework may be used for all kinds of enrichment functions and has been used in a variety of applications. As the enrichment represents near-wall turbulent fluctuations in an averaged sense and large eddies in the bulk of the flow are resolved, the method suggested may be interpreted as a detached-eddy simulation. In this respect, the authors are not seeking to point out the limitations of DES but to pioneer a new class of wall models likely to engage many ideas of previous wall modeling approaches in future developments.

The method has been validated with the two most important flow regimes for wall-bounded turbulent flow, that is an attached boundary layer represented by turbulent channel flow and separated flow featuring a strong adverse pressure gradient present in flow past periodic hills and a backward-facing step. Turbulent channel flow has been evaluated for several Reynolds numbers giving rise to the following conclusions: The method enables the use of extremely coarse meshes where barely the largest eddies are resolved while high quality of the mean velocity is guaranteed. The results for the mean velocity profile have also shown to be essentially independent of the mesh employed. Root-mean-square values of the velocity fluctuations require slightly finer resolutions, however, which is due to the fact that the function space is "tuned" to represent the mean velocity and not the fluctuations. Flow past periodic constrictions and backward-facing-step flow exhibits the large potential of the presented method for many practical applications with high pressure gradient and under separated flow conditions. While standard wall models fail to predict the correct wall shear stresses, the present enrichment-based wall model predicts the skin-friction and pressure coefficients including separation and reattachment points accurately even with very coarse meshes.

% If in two-column mode, this environment will change to single-column format so that long equations can be displayed. 
%\begin{widetext}
%$$\mbox{put long equation here}$$
%\end{widetext}

% Figures should be put into the text as floats. 
% Use the graphics or graphicx packages (distributed with LaTeX2e).
% Put the label that you will use with \ref{} command in the braces of the \label{} command.
%
% \begin{figure}
% \end{figure}

% \begin{table}
% \end{tabular}
% \end{table}

\section*{Acknowledgments}
 Computational resources provided by the Leibniz Supercomputing Centre under the project pr83te are gratefully acknowledged. The authors thank Ursula Rasthofer for helpful and inspiring discussions on the turbulence modeling approach and Stephan J\"ager for preparations towards the periodic hill benchmark.
 
\appendix
\section{Derivatives of the enrichment in cartesian coordinates} 
\label{sec:appendix}
The first and second derivatives of the enrichment with respect to cartesian coordinates, which are required for evaluation of the Galerkin formulation (\ref{eq:semidiscret}), are obtained by applying the chain rule iteratively starting from equation (\ref{eq:enrichment}). The equations are split into three groups: (i.) expressions in cartesian coordinates, (ii.) transformation to the wall coordinate $y^+$  and (iii.) derivative with respect to $y^+$. As usual, indices define space dimensions $i,j \in \{1,2,3\}$.

\renewcommand{\theenumi}{\roman{enumi}}
\begin{enumerate}
\item The first derivative is
\begin{align}
& \frac{\partial}{\partial x_i}\tilde{\bm{u}}^h(\bm{x},t)  = \nonumber \\
& \sum_{B \in N_{enr}^{\bm{u}}} \Big( \frac{\partial N_{B}^{\bm{u}}(\bm{x})}{\partial x_i}  (\psi(\bm{x},t)-\psi(\bm{x}_{B},t)) r^h(\bm{x})
 + N_{B}^{\bm{u}}(\bm{x}) \frac{\partial \psi(\bm{x},t)}{\partial x_i}  r^h(\bm{x}) 
 + N_{B}^{\bm{u}}(\bm{x}) (\psi(\bm{x},t)-\psi(\bm{x}_{B},t)) \frac{\partial r^h(\bm{x})}{\partial x_i}  \Big) \tilde{\bm{u}}_{B}
\end{align}
and the second derivative gives
\begin{align}
& \frac{\partial^2}{\partial x_i x_j}\tilde{\bm{u}}^h(\bm{x},t)= \nonumber \\
& \sum_{B \in N_{enr}^{\bm{u}}} \Big( \frac{\partial^2 N_{B}^{\bm{u}}(\bm{x})}
{\partial x_i \partial x_j}  (\psi(\bm{x},t)-\psi(\bm{x}_{B},t)) r^h(\bm{x})
 + \frac{\partial N_{B}^{\bm{u}}(\bm{x})}{\partial x_i}  \frac{\partial \psi(\bm{x},t)}{\partial x_j} r^h(\bm{x}) 
 + \frac{\partial N_{B}^{\bm{u}}(\bm{x})}{\partial x_i}  (\psi(\bm{x},t)-\psi(\bm{x}_{B},t)) \frac{\partial r^h(\bm{x})}{\partial x_j} \nonumber \\
& + \frac{\partial N_{B}^{\bm{u}}(\bm{x})}{\partial x_j} \frac{\partial \psi(\bm{x},t)}{\partial x_i}  r^h(\bm{x})
 + N_{B}^{\bm{u}}(\bm{x}) \frac{\partial^2 \psi(\bm{x},t)}{\partial x_i \partial x_j}  r^h(\bm{x})
 + N_{B}^{\bm{u}}(\bm{x}) \frac{\partial \psi(\bm{x},t)}{\partial x_i}  \frac{\partial r^h(\bm{x})}{\partial x_j} \nonumber \\
& + \frac{\partial N_{B}^{\bm{u}}(\bm{x})}{\partial x_j} (\psi(\bm{x},t)-\psi(\bm{x}_{B},t)) \frac{\partial r^h(\bm{x})}{\partial x_i}
 + N_{B}^{\bm{u}}(\bm{x}) \frac{\psi(\bm{x},t)}{\partial x_j} \frac{\partial r^h(\bm{x})}{\partial x_i}
+ N_{B}^{\bm{u}}(\bm{x}) (\psi(\bm{x},t)-\psi(\bm{x}_{B},t)) \frac{\partial^2 r^h(\bm{x})}{\partial x_i \partial x_j}  \Big)\tilde{\bm{u}}_{B}.
\end{align}
The ramp function is defined node-wise and interpolated with the standard FE expansion, allowing for straight-forward computation of its derivatives.
\item As the enrichment function is defined via the wall coordinate, its derivatives are transformed to $y^+$ yielding
\begin{equation}
\frac{\partial \psi(\bm{x},t)}{\partial x_i} = \frac{d \psi}{d y^+}\frac{\partial y^+}{\partial x_i}
\end{equation}
with
\begin{equation}
\frac{\partial y^+}{\partial x_i} = \bigg( \frac{\frac{\partial y^{h}}{\partial x_i}}{\nu}\sqrt{\frac{\tau_w^{3h}}{\rho}} + \frac{y^h \frac{\partial \tau_w^{3h}}{\partial x_i}}{2 \nu \sqrt{\tau_w^{3h}\rho}} \bigg)
\end{equation}
where $\frac{\partial y^{h}}{\partial x_i}$ and $\frac{\partial \tau_w^{3h}}{\partial x_i}$ are obtained in a straight-forward manner via the standard FE expansion (\ref{eq:ydiscret}) and (\ref{eq:fbased}). Applying the chain rule successively, the second derivative becomes
\begin{equation}
\frac{\partial^2 \psi(\bm{x},t)}{\partial x_i \partial x_j} 
= \frac{d^2 \psi}{d y^{+2}}\frac{\partial y^+}{\partial x_j}\frac{\partial y^+}{\partial x_i}
+\frac{d \psi}{d y^+}\frac{\partial^2 y^+}{\partial x_i \partial x_j}
\end{equation}
with
\begin{equation}
\frac{\partial^2 y^+}{\partial x_i \partial x_j} = \bigg( \frac{\frac{\partial^2 y^{h}}{\partial x_i \partial x_j}}{\nu}\sqrt{\frac{\tau_w^{3h}}{\rho}} 
+\frac{\frac{\partial y^{h}}{\partial x_i}\frac{\partial \tau_w^{3h}}{\partial x_j}}{2 \nu \sqrt{\tau_    w^{3h}\rho}}
+  \frac{\frac{\partial y^{h}}{\partial x_j} \frac{\partial \tau_w^{3h}}{\partial x_i}}{2 \nu \sqrt{\tau_w^{3h}\rho}}
- \frac{y^h \frac{\partial \tau_w^{3h}}{\partial x_i}\frac{\partial \tau_w^{3h}}{\partial x_j}}{4 \nu \sqrt{\rho}(\tau_w^{3h})^{3/2}}
+ \frac{y^h \frac{\partial^2 \tau_w^{3h}}{\partial x_i \partial x_j}}{2 \nu \sqrt{\tau_w^{3h}\rho}} \bigg).
\end{equation}
\item The derivatives of $\psi(\bm{x},t)$ with respect to $y^+$ may be obtained explicitly with given $\psi$ as
\begin{equation}
\frac{d\psi}{d y^+} = \frac{1}{\frac{1}{\kappa}+e^{-\kappa B}(e^{\psi}-1-\psi-\frac{\psi^2}{2!}-\frac{\psi^3}{3!})}
\end{equation}
and
\begin{equation}
\frac{d^2\psi}{d y^{+2}} = -e^{-\kappa B}\Big(e^{\psi}-1-\psi-\frac{\psi^2}{2!}\Big)\Big(\frac{d\psi}{d y^+}\Big)^{3}.
\end{equation}
\end{enumerate}

\bibliography{arxiv-v1-krank-wall-xwall}

\end{document}